\def\d{\mbox{d}}
\documentclass[showpacs,twocolumn]{revtex4}
\usepackage{graphicx}

\begin{document}

\title{Dilute Fermi gas: kinetic and interaction energies}

\author{A. A. Shanenko}

\affiliation{Bogoliubov Laboratory of Theoretical Physics,
Joint Institute for Nuclear Research, 141980, Dubna, Moscow 
region,Russia}

\date{January 13, 2004}

\begin{abstract}
A dilute homogeneous 3D Fermi gas in the ground state is considered 
for the case of a repulsive pairwise interaction. The low-density 
(dilution) expansions for the kinetic and interaction energies of 
the system in question are calculated up to the third order in the 
dilution parameter. Similar to the recent results for a Bose gas, 
the calculated quantities turn out to depend on a pairwise 
interaction through the two characteristic lengths: the former, 
$a$, is the well-known s-wave scattering length, and the latter, 
$b$, is related to $a$ by $b=a-m (\partial a/\partial m)$, where 
$m$ stands for the fermion mass. To take control of the results, 
calculations are fulfilled in two independent ways. The first 
involves the Hellmann-Feynman theorem, taken in conjunction with a 
helpful variational theorem for the scattering length. This way is 
used to derive the kinetic and interaction energies from the familiar 
low-density expansion of the total system energy first found by Huang 
and Yang. The second way operates with the in-medium pair wave 
functions. It allows one to derive the quantities of interest``from 
the scratch'', with no use of the total energy. An important result
of the present investigation is that the pairwise interaction of 
fermions makes an essential contribution to their kinetic energy. 
Moreover, there is a complicated and interesting interplay of these 
quantities. 
\end{abstract}


\maketitle

%--------------------------------------------------------------
\section{Introduction}
\label{1sec}

Recent experiments with magnetically trapped alkali atoms
significantly renewed interest in properties of quantum gases. As
it is known, the initial series of these experiments concerned a 
Bose gas~($^{87}Rb$~\cite{ander}, $^{23}Na $~\cite{davis}, and 
$^{7}Li$~\cite{brad}) and resulted in extensive reconsiderations and 
new investigations in the field of the Bose-Einstein condensation. 
In so doing theoretical and experimental observations were made that
not only confirmed conclusions made more than forty years ago
but also provided a new horizon of the boson physics. In
particular, one should point out good agreement of the results of
solving the Gross-Pitaevskii equation derived in 1960s~\cite{gp}
with experimental data on the density profiles of a trapped Bose
gas~\cite{hau}. The so-called release energy measured in the
experiments with rubidium was also found to be in good agreement
with theoretical expectations based on the time-dependent
Gross-Pitaevskii equation~\cite{hol}. Among new theoretical
achievements the exact derivation of the Gross-Pitaevskii energy
functional~\cite{liebA} can be mentioned along with the proof that
a Bose gas with repulsive interaction is $100\%$ superfluid in the
dilute limit~\cite{liebB}. As to the experimental innovations, 
observations of interference of two Bose 
condensates~\cite{andrews,kett} is a good example (for 
interesting theoretical details see the papers~\cite{cirac}, 
\cite{devr} and reviews \cite{pit},\cite{legett}). 

The first communications concerning experiments with trapped fermionic 
atoms appeared in the literature about three years ago~\cite{marco} 
when a temperature near $0.4T_f$ was claimed to be reached for a 
trapped $^{6}Li$, where $T_f$ is the temperature below which the 
Fermi statistics is of importance. Nowadays the temperatures close 
to $0.2T_f$~\cite{schreck} and $0.5T_f$~\cite{hadzibabic} are
reported for the $ ^{6}Li$-vapor. Whereas atoms of fermionic
$^{40}K$ were recently cooled down to $0.3T_f$~\cite{roati}. So,
the regime of the degenerate Fermi gas is already under
experimental study. In view of this fact, reconsideration of
the basic aspects of the theory of a dilute uniform Fermi gas
in the ground state is of importance. 

In the present paper a dilute Fermi gas with repulsive pairwise 
interaction is under consideration. Why the situation of a 
repulsive Fermi gas is of interest whereas the $s-$wave scattering 
length is negative for $^{6}Li$~\cite{stoof} and, most likely, for 
$^{40}K$~\cite{roati} so that a trapped $^{6}Li$ is considered as a 
good candidate for observation of BCS-like
transition~\cite{stoof,combescot,houbiers}? The point is that the
experiments can produce (and is now producing) the temperatures at
which the BCS pairing does not occur yet. So, at these temperature 
a Fermi gas with attractive pairwise interaction is close enough in 
properties (with the corrections of $O((T/T_f)^2)$, where $T/T_f 
\approx 0.2$) to a repulsive Fermi gas in the ground state. Of course, 
with one obvious alteration: the positive $s-$wave scattering length 
should be replaced by a negative one in the final expressions~(see, 
for example, \cite{stoof}). Thus, the experiments with magnetically 
trapped atoms of $^{6}Li$ and $^{40}K$ offer exciting possibility 
of exploring the both superfluid and normal states of a Fermi gas. 
In addition, the most recent publications~\cite{ohara,bourdel} 
demonstrate that there exists interesting possibility of ruling the 
scattering length of $^{6}Li$ which varies in a wide range of values, 
from negative to positive ones, when a magnetic field is imposed on 
the system.   

The particular problem to be investigated here concerns the kinetic 
$E_{\rm kin}$ and interaction $E_{\rm int}$ energies of a uniform 
dilute 3D Fermi gas in the ground state and with a repulsive 
interparticle potential. This problem is connected with a more general 
question related to all the quantum gases. The question is if the 
pairwise interaction of quantum particles makes contribution to the 
kinetic energy of a quantum gas or not? It is well-known that for a 
classical imperfect gas the pairwise interaction does not make any
contribution to the kinetic energy. The usual expectations regarding 
the kinetic energy of dilute quantum gases comes from the 
pseudopotential approach. According to these expectations the kinetic 
energy of a dilute quantum gas is not practically affected by the 
pairwise interaction. It means that taken in the leading order of 
the expansion in the dilution parameter, the total system energy 
of a ground-state Bose gas coinsides with the interaction one 
if calculated with the pseudopotential~(see 
Refs.~\cite{pit,shanA,shanB}). For the Fermi case the same approach 
dictates that the kinetic energy does not include terms depending 
on the pairwise potential in the leading and next-to-leading orders 
of the dilution expansion (see Ref.~\cite{vichi} and Eqs. 
(\ref{ekinPS}) and (\ref{eintPS}) below). In other words, in what 
concerns relation between the kinetic and interaction energies, a 
quantum gas is very similar to a classical one from the pseudopotential 
viewpoint. However, this result was proved to be wrong. An adequate 
and thorough procedure of calculating $E_{\rm kin}$ and $E_{\rm int}$ 
of a cold dilute Bose gas has been recently developed in 
Refs.~\cite{shanA,shanB}. It proves that the pairwise interaction in a 
Bose gas has a strong effect on the kinetic energy. Moreover, there are 
quite real situations when the kinetic energy of a uniform dilute Bose 
gas is essentially more than the interaction one! It is now necessary 
to clarify this situation in the Fermi case. The more so, that the 
interaction and kinetic energies of imperfect trapped quantum 
gases are now under experimental study~\cite{vichi,bourdel}. Thus, 
the aim of the present publication is to generalize the procedure 
developed in \cite{shanA,shanB} to the Fermi case.

The paper is organized as follows. The Section \ref{sec2} presents
the kinetic and interaction energy of a ground-state repulsive
Fermi gas found with the Hellmann-Feynman theorem on the basis of
an auxiliary variational relation given in Refs.~\cite{shanA,shanB}. 
The Section \ref{sec3} is to consider the derived expressions in 
various regimes: from a weak coupling to a strong one. This is needed 
to discuss the failure of the pseudopotential approach in operating 
with $E_{\rm kin}$ and $E_{\rm int}$. Derivation of $E_{\rm kin}$ and 
$E_{\rm int}$ in Section 2 is simple but rather formal so that some 
questions can remain.  This is why Sections \ref{sec4} and \ref{sec5}
give a more physically sound way of calculating the kinetic and 
interaction energies. This way invokes a method developed in the 
papers~\cite{shanA,shanB} and dealing with the pair wave functions, 
which allows one to go in more detail concerning the microscopic 
features of dilute quantum gases.
%%%%%%%%%%%%%%%%%%%%%%%%%%%%%%%%%%%%%%%%%%%%%%%%%%%%%%%%%%%%%%%%%%%%%%
\section{Hellmann-Feynman theorem}
\label{sec2}

Let us consider the system of $N$ identical fermions placed in a
box with the volume $V$ and ruled by the following Hamiltonian:
\begin{equation}
\hat H=-\sum_{i}\frac{\hbar^2}{2m}\nabla^2_{i}+\sum_{i>j}
V(|{\bf r}_i-{\bf r}_j|) \label{H}
\end{equation}
with the pairwise interaction $V(r)=\gamma \Phi(r)$, where $\gamma>0$ 
is the coupling constant and $\Phi(r)>0$ stands for the interaction 
kernel~($r=|{\bf r}|$). Below the particle spin is assumed to be 
$s=1/2$~\cite{note1}. The ground-state energy of the system in 
question $E=\langle0|\hat{H}|0\rangle$ obeys the well-known relation
\begin{equation}
\delta E=\langle 0|\delta\hat{H}|0 \rangle
\label{HF}
\end{equation}
called the Hellmann-Feynman theorem, $\delta E$ and $\delta
\hat{H}$ being infinitesimal changes of $E$ and $\hat{H}$. An
advantage of this theorem is that it yields important relations
connecting the total ground-state energy $E$ with the kinetic
$E_{\rm kin}$ and interaction $E_{\rm int}$ energies. These
relations read
\begin{eqnarray}
-m\frac{\partial E}{\partial m}=\Bigl\langle0\Bigl|-\sum_{i}
\frac{\hbar^2}{2m}\nabla^2_{i}\Bigr|0\Bigr\rangle
=E_{\rm kin} \label{ekin},\\
\gamma \frac{\partial E}{\partial \gamma}=\Bigl\langle0\Bigl|
\sum_{i>j}V(|{\bf r}_i-{\bf r}_j|)\Bigr|0\Bigr\rangle
=E_{\rm int}.\label{eint}
\end{eqnarray}
If the dependence of the ground-state energy on the coupling
constant and particle mass were known explicitly, one would
readily be able to calculate $E_{\rm kin}$ and $E_{\rm int}$ by
means of Eqs.~(\ref{ekin}) and (\ref{eint}). However, it is not
the case as a rule, and the dependence is usually given only
implicitly.

In the situation of the repulsive Fermi gas the dependence of the
ground-state energy on $\gamma$ and $m$ is indeed known only
implicitly. According to the familiar result of Huang and
Yang~\cite{hy}  found with the pseudopotential approach but then 
reproduced within the boundary collision expansion method~\cite{ly}
beyond any effective-interaction arguments, the energy per fermion
$\varepsilon=E/N$ reads
\begin{equation}
\varepsilon=\frac{3\hbar^2 k^2_F}{10m}\biggl[1+\frac{10}{9\pi}k_Fa
+\frac{4}{21\pi^2}(11-2\ln2)k^2_Fa^2\biggr],
\label{HY}
\end{equation}
which is accurate to the terms of order $k^2_Fa^2$. In Eq.~(\ref{HY}) 
$a$ stands for the $s-$wave scattering length, $k_F$ is the Fermi 
wavenumber given by
\begin{equation}
k_F=(3\pi^2 n)^{1/3},
\label{pf}
\end{equation}
where $n=N/V$, and the thermodynamic limit $N \to \infty, \;V \to
\infty, \;n=N/V \to const$ is implied. Inserting Eq.~(\ref{HY}) 
in Eqs.~(\ref{ekin}) and (\ref{eint}), one can arrive at
{\setlength\arraycolsep{0pt}
\begin{eqnarray}
&&\varepsilon_{\rm kin}=\nonumber\\
&&=\varepsilon-\frac{3\hbar^2 k^2_F}{10a}\,
\frac{\partial a}{\partial m} \biggl[\frac{10}{9\pi}k_Fa
+\frac{8}{21\pi^2}(11\!-\!2\ln2)k^2_Fa^2\biggr],\quad\label{ekin1}\\
&&\varepsilon_{\rm int}\!=\!\frac{3\hbar^2 k^2_F}{10ma}\,\gamma
\frac{\partial a}{\partial \gamma} \biggl[\frac{10}{9\pi}k_Fa
+\frac{8}{21\pi^2}(11\!-\!2\ln2)k^2_Fa^2\biggr],\quad\label{eint1}
\end{eqnarray}}where $\varepsilon_{\rm kin}=E_{\rm kin}/N$ and 
$\varepsilon_{\rm int}=E_{\rm int}/N$. Hence, to derive the kinetic 
and interaction energies from Eq.~(\ref{HY}) with the help of the 
Hellmann-Feynman theorem, we should have an idea concerning the 
derivatives of $a$ with respect to the particle mass $m$ and coupling 
constant $\gamma$. As $E=E_{\rm kin}+E_{\rm int}$, then from 
Eqs.~(\ref{ekin1}) and (\ref{eint1}) it follows that
\begin{equation}
m\,\partial a/\partial m
=\gamma\,\partial a/\partial \gamma.
\label{mgam}
\end{equation}
This property of the derivatives becomes clear if we remind that in 
the $3D$ case the $s-$wave scattering length is given by
\begin{equation}
a=\frac{m\gamma}{4\pi \hbar^2}\int d^3r \varphi(r) \Phi(r),
\label{aA}
\end{equation}
where $\varphi(r)$ obeys the the two-body Schr\"odinger equation in 
the center-off-mass system:
\begin{equation}
-(\hbar^2/m\gamma)\nabla^2\varphi(r)+\Phi(r)\varphi(r)=0.
\label{schr}
\end{equation}
The pair wave function $\varphi(r)$ represents the
ze\-ro-mo\-men\-tum scattering state, and $\varphi(r) \to 1$ when 
$r \to \infty$. The scattering part of the pair wave function given
by the definition $\varphi(r)=1+\psi(r)$ is specified by the
following asymptotic behavior:
\begin{equation}
\psi(r) \to -a/r \quad (r \to \infty).
\label{as}
\end{equation}
Note once more that the pairwise potential involved in 
Eqs.~(\ref{aA}) and (\ref{schr}) is $V(r)=\gamma \Phi(r)$ but not 
$\Phi(r)$ which is the repulsive interaction kernel. As it is seen 
from Eqs.~(\ref{aA}) and (\ref{schr}), the scattering length depends 
on the particle mass and coupling constant through the product
$m\gamma$. Hence, to use Eqs.~(\ref{ekin1}) and (\ref{eint1}) we
should know the derivative of $a$ with respect to $m\gamma$.

This derivative can be found with very useful variational theorem
proved in the papers~\cite{shanA,shanB}. After small algebra the
result of this theorem is rewritten in the following form:
\begin{equation}
\delta a=\frac{\delta (m\gamma)}{4\pi \hbar^2} \int d^3r
\varphi^2(r) \Phi(r),
\label{theorem}
\end{equation}
where, remind, $\varphi(r)$ is a real quantity. In view of crucial
importance of this theorem, let us make an explaining remark
concerning the proof. The key point here is to represent
Eq.~(\ref{aA}) as
\begin{equation}
a=\frac{m\gamma}{4\pi \hbar^2}\int d^3r \varphi^2(r)\Phi(r)+
\frac{1}{4\pi} \int d^3 r |\nabla \psi(r)|^2,
\label{aB}
\end{equation}
which is realized with the help of Eqs.~(\ref{schr}), (\ref{as}) and
$\nabla(\psi \nabla\psi)=\nabla \psi\;\nabla \psi+\psi\nabla^2\psi$.
So, from Eq.~(\ref{theorem}) one gets
\begin{equation}
m \partial a/\partial m=\gamma \partial a/\partial \gamma=a-b,
\label{der}
\end{equation}
where the additional characteristic length $b>0$ is of the form
\begin{equation}
b=\frac{1}{4\pi}\int d^3r |\nabla \psi(r)|^2.
\label{bA}
\end{equation}
Emphasize that $b$ can not be represented as a function of $a$ in
principle, and the ratio $b/a$ depends on a particular shape of a
pairwise potential involved. Now we need nothing more to calculate
the kinetic and interaction energies of the uniform repulsive Fermi
gas in the ground state. Equations~(\ref{ekin1}) and (\ref{eint1})
taken in conjunction with Eq.~(\ref{der}), result in the following
expressions:
\begin{eqnarray}
\varepsilon_{\rm kin}\!=&&\!\frac{3\hbar^2 k^2_F}{10m}\biggl[
1+\frac{10}{9\pi}k_Fb\nonumber\\
&&+\frac{4}{21\pi^2}(11\!-\!2\ln2)\left(2\frac{b}{a}-1\right)
k^2_Fa^2\biggr],
\label{ekin2}\\
\varepsilon_{\rm int}\!=&&\!\frac{3\hbar^2 k^2_F}{10m}
\left(1-\frac{b}{a}\right)\! \nonumber\\
&&\times\biggl[\frac{10}{9\pi}k_Fa
+\frac{8}{21\pi^2}(11\!-\!2\ln2)k^2_Fa^2\biggr],
\label{eint2}
\end{eqnarray}
whose sum is, of course, equal to Eq.~(\ref{HY}). We again have series
expansions in $k_Fa$ but with coefficients depending on the ratio 
$b/a$.
%%%%%%%%%%%%%%%%%%%%%%%%%%%%%%%%%%%%%%%%%%%%%%%%%%%%%%%%%%%%%%%%%%%%%%%%
\section{From weak to strong coupling}
\label{sec3}

To go in more detail concerning  Eqs.~(\ref{ekin2}) and
(\ref{eint2}), let us consider them in various regimes. We speak
about the week coupling when the interaction kernel $\Phi(r)$
is integrable and the coupling constant $\gamma \ll 1$. The
integrable kernel with $\gamma \gg 1$ and a singular pairwise
interaction like the hard-sphere potential are related to the
strong-coupling regime. The expansion parameter $k_Fa$ involved
in the expressions mentioned above corresponds to the dilution
limit $k_F \to 0$. In this situation one is able to operate with
Eq.~(\ref{HY}) in the both weak- and strong-coupling cases. However,
for the weak coupling $k_Fa$ is small even beyond the dilute regime
due to $a \propto \gamma \ll 1$. This is why Eq.~(\ref{HY}) can be
used and rearranged in such a way that to derive the weak-coupling
expansion for $\varepsilon$.

In the weak-coupling regime the scattering length $a$ is given by the
Born series:
\begin{equation}
a =a_{0}+a_{1}+\ldots
\label{Bser}
\end{equation}
with
\begin{equation}
a_{0}=\frac{m\gamma}{4\pi\hbar^{2}}\Phi(k=0),\;
a_{1\!}=\!-\frac{m\gamma^2}{4\pi\hbar^{2}}\int
\frac{\d^{3}k}{(2\pi)^{3}}\frac{\Phi^{2}(k)}{2T_{k}}, \label{aBorn}
\end{equation}
where $T_k=\hbar^2 k^2/(2m)$, and $\Phi(k)$ is the Fourier transform
of the interaction kernel ( for more detail see Ref.~\cite{brue}). 
Inser\-ting Eq.~(\ref{Bser}) in Eq.~(\ref{HY}), one gets the 
fol\-low\-ing expression:
{\setlength\arraycolsep{0pt}
\begin{eqnarray}
\varepsilon=&&\frac{3\hbar^2 k^2_F}{10m}\Biggl[1+\frac{10}{9\pi}k_F
a_0\nonumber\\
&&+\biggl(\frac{10}{9\pi}k_Fa_1+\frac{4}{21\pi^2}(11\!-\!2\ln2)
k^2_Fa_0^2\biggr)\Biggr],
\label{LYw}
\end{eqnarray}
where terms of order $\gamma^3$ are ignored. Due to Eq.~(\ref{aBorn}) 
the dependence of Eq.~(\ref{LYw}) on the particle mass and coupling
constant is known explicitly. Hence, one can readily employ the
Hellmann-Feynman theorem that, taken together with Eq.~(\ref{LYw}),
yields
{\setlength\arraycolsep{0pt}
\begin{eqnarray}
\varepsilon_{\rm kin}=&&\frac{3\hbar^2 k^2_F}{10m}\Biggl[1-\nonumber\\
&&-\biggl(\frac{10}{9\pi}k_Fa_1+\frac{4}{21\pi^2}(11\!-\!2\ln2)
k^2_Fa_0^2\biggr)\Biggr],
\label{ekinW}
\end{eqnarray}}
{\setlength\arraycolsep{0pt}
\begin{eqnarray}
\varepsilon_{\rm int}=&&\frac{3\hbar^2 k^2_F}{10m}\Biggl[
\frac{10}{9\pi}k_Fa_0\nonumber\\
&&+\biggl(\frac{20}{9\pi}k_Fa_1
+\frac{8}{21\pi^2}(11\!-\!2\ln2)
k^2_Fa_0^2\biggr)\Biggr].
\label{eintW}
\end{eqnarray}}
So, the derived results suggest that the pairwise interaction
influences the both kinetic and interaction energies of a Fermi
gas. In the weak-coupling regime the major part of the
$\gamma$-dependent contribution to Eq.~(\ref{LYw}) is related to
$E_{\rm int}$, this part being proportional to $\gamma$. While the
terms of order $\gamma^2$ appear in both $E_{\rm kin}$ and $E_{\rm
int}$. This conclusion meets usual expectations according to which
the contribution to the mean energy coming from the pairwise
potential is mostly the interaction energy for dilute quantum 
gases~(see, for example, Refs.~\cite{pit,stoof} and the discussion in 
Introduction of the paper~\cite{shanA}). On the contrary, beyond
the weak-coupling regime the situation with $E_{\rm kin}$ and
$E_{\rm int}$ turned out to be rather curious and differs
significantly from that of the weak-coupling case. However, before
any detail let us discuss the pseudopotential predictions for
$E_{\rm kin}$ and $E_{\rm int}$ being the basis of usual
speculations involving the kinetic and interaction energies of
quantum gases beyond the weak-coupling regime.

At present the customary way of operating with the thermodynamics 
of a  dilute cold Fermi gas with repulsive pairwise potential is based 
on the effective-interaction procedure: one is able to use either the 
$t-$matrix formulation like in the Galitskii original paper~\cite{gal} 
or the pseudopotential scheme applied in the classical work of Huang 
and Yang~\cite{hy}. In the dilution limit the $t-$matrix  is reduced 
to $t=4\pi\hbar^2a/m$, which yields the momentum-independent result
$4\pi\hbar^2a/m$ for the Fourier transform of the effective
interaction. This is why one is able not to make essential difference
between these two effective-interaction formulations both referred to 
as the pseudopotential approach here. The key point of
this approach is that to go beyond the weak-coupling regime, one
should replace the Fourier transform of the pairwise interaction
$\Phi(k)$ by the quantity $4\pi\hbar^2a/m$ in all the expressions
related to the weak-coupling approximation. In so doing, some
divergent integrals  appear due to ignorance of the momentum
dependence of the $t-$matrix. Indeed, substituting $t=4\pi 
\hbar^2 a/m$ for $\Phi(k)$ in Eq.~(\ref{aB}), one gets a divergent 
quantity $a_1\propto \int\d^{3}k/2T_{k}$ that makes contribution to
the total energy of the system. To derive the classical result of 
Huang and Yang, the divergent term proportional to $a_1$ should be 
removed, which is usually fulfilled via a regularization procedure
simulating the momentum dependence of the $t-$matrix.  In the 
pseudopotential scheme of Huang and Yang this corresponds to use
of the effective interaction $(4\pi\hbar^2a/m)\delta(r)(\partial 
/\partial r) r$ rather than $(4\pi\hbar^2a/m)\delta(r)$.  From this
one can learn that to generalize Eqs.~(\ref{LYw}), (\ref{ekinW})
and (\ref{eintW}) to the situation of a finite coupling constant,
one should replace $a_0$ by $a$ and remove all the terms depending
on $a_1$ in the mentioned equations. This yields Eq.~(\ref{HY}) 
and the following pseudopotential predictions for the kinetic and 
interaction energies:
{\setlength\arraycolsep{0pt}
\begin{eqnarray}
\varepsilon^{(\rm ps)}_{\rm kin}=\frac{3\hbar^2 k^2_F}{10m}
\biggl[1-\frac{4}{21\pi^2}(11\!-\!2\ln2) k^2_Fa^2\biggr]&& ,\quad
\label{ekinPS}\\
\varepsilon^{(\rm ps)}_{\rm int}=\frac{3\hbar^2
k^2_F}{10m}\biggl[ \frac{10}{9\pi}k_Fa
+\frac{8}{21\pi^2}(11\!-\!2\ln2) k^2_Fa^2\biggr]&&. \quad
\label{eintPS}
\end{eqnarray}}Note that these results can be derived in 
another way as well. For example, the first term in 
Eq.~(\ref{eintPS}) can readily be reproduced with the 
pseudopotential in the Hartree-Fock approximation~(see 
Ref.~\cite{hy} and the next section of the present paper). From 
Eqs.~(\ref{ekinPS}) and (\ref{eintPS}) one could conclude that 
the second term in Eq.~(\ref{HY}) is related to
the interaction energy, and, hence, the contribution to the mean 
energy of a dilute cold Fermi gas coming from the pairwise 
potential is mainly the interaction energy. However, now we know 
that actually it is not the case. So, one should be careful with the 
pseudopotential procedure which has serious limitations in spite of
the correct result for the mean energy. Here it is worth noting that
the pseudopotential scheme preserves some features of the 
weak-coupling regime even being applied in the strong-coupling 
case. This concerns the relation between the kinetic and interaction 
energy for both a dilute Fermi ground-state gas and a Bose 
one~\cite{shanA,shanB}. The same problem appears when the 
pseudopotential is used to calculate the two-particle Green function
in a Bose gas, which manifests itself in abnormal short-range boson 
correlations~\cite{shanA}. Similar troubles can also be expected for 
the two-fermion Green function.

Now let us consider Eqs.~(\ref{ekin2}) and (\ref{eint2}) beyond the 
weak-coupling regime, the ratio $b/a$ being of special interest. We
start with the simplified situation of penetrate-able spheres that are
specified by the interaction kernel
\begin{equation}
\Phi(r)=\left\{\begin{array}{rr}
\Phi\quad      & {\rm if}\quad r\leq r_0,\\[1mm]
0\quad         & {\rm if} \quad r > r_0.
\end{array}\right.
\label{ker_sph} 
\end{equation}
Inserting Eq.~(\ref{ker_sph}) in Eq.~(\ref{schr}), one can find
\begin{equation}
\varphi(r)=\left\{\begin{array}{rr}
2A\sinh(\alpha r)/r \quad  & {\rm if} \quad r\leq r_0,\\[1mm]
1-a/r \quad                & {\rm if} \quad r> r_0,
\end{array}\right.
\label{phisph} 
\end{equation}
where $\alpha^2=m\gamma \Phi/\hbar^2\;(\Phi > 0)$ and $A$ is 
a constant. Equation~(\ref{phisph}) taken together with the usual 
boundary conditions at $ r=r_0$ leads to 
\begin{equation}
a=r_0\Bigl[ 1-\tanh(\alpha r_0)/(\alpha r_0)\Bigr]
\label{aC}
\end{equation}
and
\begin{equation}
b\!=\!r_0\!\left[1\!-\!\frac{1}{2}\Bigl(3\tanh(\alpha r_0)/(\alpha r_0)
\!-\!{\rm csch}(\alpha r_0)\Bigr)\right]\!,
\label{bB}
\end{equation}
where  ${\rm csch}(x)=1/\cosh^2(x)$. One can readily check that in the 
weak-coupling regime, when $\alpha r_0 \propto \gamma^{1/2}\to 0$,  
Eqs.~(\ref{aC}) and (\ref{bB}) are reduced to
\begin{equation}
a\simeq\frac{1}{3}\alpha^2r_0^3\propto \gamma, \quad
b\simeq \frac{2}{15}\alpha^4r_0^5\propto \gamma^2
\label{sphW}
\end{equation}
and, hence, $b\ll a$. This means that the next-to-leading term in the 
expansion in $k_Fa$ given by Eq.~(\ref{HY}) is mostly the interaction 
energy, as it was mentioned above. On the contrary, in the 
strong-coupling regime,  when $\alpha r_0 \to \infty$, Eqs.~(\ref{aC}) 
and (\ref{bB}) give
\begin{equation}
a \to r_0, \quad b \to r_0.
\label{sphS}
\end{equation}   
Hence,  $b/a\to 1$, and the ground-state energy of a dilute Fermi 
gas with the hard-sphere interaction is exactly kinetic!  Note that the 
same conclusion is valid for a dilute Bose gas of the hard 
spheres~\cite{shanA,shanB,liebC}.

Another, a more realistic example concerns a situation when 
the interaction kernel combines a short-range repulsive sector with a 
long-range attractive one. Here we are especially interested in a 
negative scattering length. It is usually considered~(see, e.g., 
\cite{gora}) that for alkali atoms one can employ the following
approximation:
\begin{equation}
\Phi(r)=\left\{\begin{array}{cc}
+\infty\quad &{\rm if}\quad r \leq  r_0,\\[1mm]
-C/r^6 \quad &{\rm if}\quad r > r_0.
\end{array}\right.
\label{ker_att}
\end{equation}
The scattering length for the pair interaction kernel (\ref{ker_att}) 
is of the form~(see Ref~\cite{ties}) 
\begin{equation}
a/r_c=\Gamma(3/4)J_{-1/4}(x)/\bigl[2\Gamma(5/4)J_{1/4}(x)\bigr],
\label{aD}
\end{equation}
where $x\!=\!r_c^2/(2r_0^2),\;r_c\!=\!(m\gamma C/
\hbar^2)^{1/4}$, whereas $J_{\nu}(x)$ and $\Gamma(z)$ de\-note 
the Bessel function and the Euler gamma-function. It is known that 
$J_{\nu}(x)\simeq x^{\nu}/[2^{\nu}\Gamma(1+\nu)]$ for $x\to 0$. 
Therefore, Eq.~(\ref{aD}) reduces to $a=r_0$ in this limit. In other 
words, when the attractive sector is ``switched off'', we arrive at 
the hard-sphere result discussed in the previous paragraph 
of the present section. For $x>0$ the scattering length 
(\ref{aD}) is a decreasing function of $x$ with the 
complicated pattern of behaviour specified by the infinite set 
of singular points $\{x^{(1)}_{\infty},\;x^{(2)}_{\infty},
\;x^{(3)}_{\infty},\ldots\}$. These points are the zeros of $J_{1/4}(x)$ 
so that $a\to -\infty$ when $x\to x^{(i)}_{\infty}-0$ and $a \to 
+\infty$ when $x \to x^{(i)}_{\infty}+0$. In addition, there is also the 
infinite sequence of the zeros of the scattering length $\{x^{(1)}_{0},\;
x^{(2)}_{0},\;x^{(3)}_{0},\ldots\}$ being the zeros of $J_{-1/4}(x)$. 
Note that $x^{(i)}_{0}< x^{(i)}_{\infty}<x^{(i+1)}_{0}$. Keeping in mind 
this information and Eq.~(\ref{aD}), we can explore the ratio $b/a$ for
the pair interaction kernel (\ref{ker_att}). Equation~(\ref{aD}) leads 
to 
\begin{eqnarray}
\gamma\partial a/\partial\gamma= 
D\sqrt{x}\,\Bigl[J_{-1/4}(x)&&/\bigl(2J_{1/4}(x)\bigr)\nonumber\\
&&-\,\sqrt{2}/\bigl(\pi J^2_{1/4}(x)\bigr)\Bigr].
\label{mgam1}
\end{eqnarray}
with $D=r_0\Gamma(3/4)/[2^{3/2}\Gamma(5/4)]$. Note that to 
derive Eq.~(\ref{mgam1}), the useful formula
$$
J_{\nu+1}(x)J_{-\nu}(x)+J_{\nu}(x)J_{-(\nu+1)}(x)\!=\!
-\,2\sin(\pi \nu)/(\pi x)
$$
should be applied. Equation~(\ref{mgam1}), taken in conjunction 
with Eq.~(\ref{der}), yields 
\begin{equation}
b/a=3/4+1/\bigl[\pi \sqrt{2} J_{1/4}(x)J_{-1/4}(x)\bigr].
\label{bC}
\end{equation}
\begin{figure}[t]
\centerline{\includegraphics[width=7.cm,clip=true]{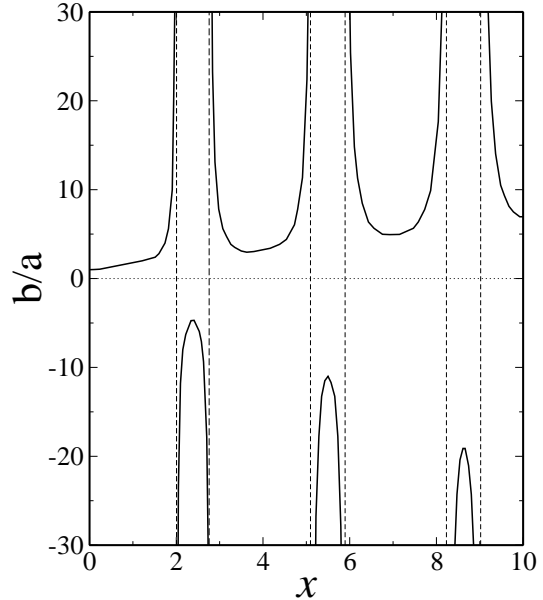}}
\caption{The ratio $b/a$ versus $x=r_c^2/(2r_0^2)$ for the pairwise 
interaction kernel (\ref{ker_att}), $r_c=(m\gamma C/\hbar^2)^{1/4}.$}
\label{fig1}
\end{figure}As it is seen from Eq.~(\ref{bC}), in the limit $x \to 0$  
we get the hard-sphere result $b/a=1$~(see Fig.~\ref{fig1}). The 
quantity $b$~(remind that $b$ is always positive!) is finite at 
$x=x^{(i)}_{0}$, while $b\to+\infty$ for $x\to x^{(i)}_{\infty}$. In 
the latter situation $b$ goes to infinity in such a way that $b/|a|
\to +\infty$ though $|a| \to \infty$ for $x\to x^{(i)}_{\infty}$, too. 
Hence, $x^{(i)}_{0}$ and $x^{(i)}_{\infty}$ are both singular points 
of $b/a$. Let us stress that the zeros of the scattering length in the 
case considered have nothing to do with the weak-coupling regime for 
which, remind, $b/|a| \ll 1 $. Operating with the kernel 
(\ref{ker_att}) we are not able to reach the weak-coupling regime at 
all because in this kernel is not bound from above. Now let us consider
the situation of a negative scattering length being of special 
interest in the experimental context. The scattering length given by 
Eq.~(\ref{aD}) is negative provided that $x^{(i)}_{0}<x
<x^{(i)}_{\infty}$. As it seen from Fig.~(\ref{fig1}), for any of these 
intervals the ratio $b/a$ has a maximum value $[b/a]^{(i)}_{\rm max}$, 
and it decreases while $i$ increases. In particular, $[b/a]^{(1)}_{\rm 
max}\approx -5$, whereas $[b/a]^{(2)}_{\rm max}\approx -12$ and 
$[b/a]^{(3)}_{\rm max} \approx -18$~(see Fig.~(\ref{fig1})). Hence,
Eq.~(\ref{bC}) turned out to make it possible to get some information 
about $b/a$ even without specifying the range of the relevant values 
of $x$~(in spite of the fact that this range 
is in principle known). Indeed, according to the mentioned above, the 
ratio $b/a$ does not exceed $[b/a]^{(1)}_{\rm max}\approx -5$ if the 
scattering length is negative. This suggests that the contribution of 
the pairwise potential to the kinetic energy is much larger than the 
absolute value of the corresponding contribution to the mean energy for 
a normal-state dilute Fermi gas with a negative scattering length at 
temperatures close enough to zero! The interaction energy is negative 
in this case and also much larger, if taken in absolute value, than the 
sum of the $a-$dependent terms in the Huang-Yang result. Note that for 
alkali atoms~\cite{ties} one can expect that $x\sim10$, which means 
that $b/|a|\gtrsim 20$~(see Fig.~\ref{fig1}). In view of the recent 
results~\cite{ohara} on a trapped Fermi gas, it is also of interest to 
consider behaviour of $b/a$ in vicinities of the special points 
$x^{(i)}_{\infty}$ and $x^{(i)}_{0}$. Varying the magnetic field acting 
on the system of $^{6}Li-$atoms, the authors of the paper~\cite{ohara} 
were interested in the regime of the Feshbach resonance, for which 
$a \to \infty$, and in the situation when $a\to 0$, as well. The 
both variants, as it follows from our consideration, are characterized 
by $b/|a| \gg 1$. Note that passage to the limit $a \to \infty$ in 
Eqs.~(\ref{ekin2}) and (\ref{eint2}) is not correct because it violates
the expansion condition $k_F|a| \ll 1$. On the contrary, one can set 
$a=0$ in these equation, which leads to the exact result, beyond the
perturbation theory,
{\setlength\arraycolsep{0pt}
\begin{eqnarray}
&&\varepsilon_{\rm kin}\bigl(a\to 0\bigr)=\frac{3\hbar^2k_F^2}{10m}+
\frac{\hbar^2 k_F^3 b}{3\pi m},\label{a=0kin}\\
&&\varepsilon_{\rm int}\bigl(a\to 0\bigr)=-\frac{\hbar^2 
k_F^3 b}{3\pi m}.
\label{a=0int}
\end{eqnarray}}Equations~(\ref{a=0kin}) and (\ref{a=0int}) correspond  
to an unusual and extreme situation that, nevertheless, is 
experimentally attainable now~(see Ref.~\cite{bourdel,ohara}). The 
total energy of the system is here equal (or practically equal) to 
that of an ideal Fermi gas, while the interaction and kinetic energies 
taken separately have nothing to do with those of a gas of 
noninteracting fermions. The most interesting particular case concerns 
the regime $k_F b \gg 1$, where the first term in Eq.~(\ref{a=0kin}) 
is negligible as compared to the second one. In this case 
$|\varepsilon_{\rm int}|\approx \varepsilon_{\rm kin} \propto n$. 

Thus, the examples listed above show that in physically relevant 
situations the correct results for the kinetic and interaction energies 
of a dilute Fermi gas given by Eqs.~(\ref{ekin2}) and (\ref{eint2}) 
differ significantly (more than by order of magnitude!) from the 
pseudopotential predictions~(\ref{ekinPS}) and (\ref{eintPS}).
As it is seen, in a Fermi gas the pairwise interaction has a profound 
effect on the kinetic energy contrary to a classical imperfect gas. 
And this well meets the conclusion on an interacting Bose gas derived 
in Refs.~\cite{shanA,shanB}. A physically sound way of explaining 
this feature of quantum gases is to invoke the formalism of the 
in-medium pair wave functions. This is why below, in Sections 
(\ref{sec4}) and (\ref{sec5}), the interaction and kinetic 
energies of a Fermi gas are investigated through the prism of this
formalism.

%%%%%%%%%%%%%%%%%%%%%%%%%%%%%%%%%%%%%%%%%%%%%%%%%%%%%%%%%%%%%%%%%%%%%%
\section{Interaction energy via the pair wave functions}
\label{sec4}

The derivation of the kinetic and interaction energies of a dilute
ground-state Fermi gas given in Section~\ref{sec2} is mathematically 
adequate. However, from the physical point of view it has an obvious 
disadvantage. Namely, the microscopic information remains hidden in 
Eqs.~(\ref{HY}), (\ref{ekin2}) and (\ref{eint2}) due to its implicit 
usage in Section \ref{sec2}. To eliminate this shortcoming, 
$\varepsilon_{\rm int}$ and $\varepsilon_{\rm kin}$ are considered 
below with a physically sound approach based on the in-medium pair 
wave functions (PWF).  

For the sake of convenience, let us begin with the interaction energy. 
It is well-known that all the microscopic information concerning the 
$N-$particle system is contained in the $N-$particle density matrix.
In the case of interest the $N-$matrix is defined by 
{\setlength\arraycolsep{0pt}
\begin{eqnarray}
\varrho_N(x_1',x_2'&&, \dots,x_N';x_1,x_2,\dots,x_N)=\nonumber\\
=&&\Psi^{*}(x_1,x_2,\dots,x_N)\Psi(x_1',x_2', \dots ,x_N'),
\label{Nm}
\end{eqnarray}}where $\Psi(x_1,x_2,\dots,x_N)$ is the ground-state 
normalized wave function, $\;x=\{{\bf r},\sigma\}$ stands for the 
space coordinates ${\bf r}$ and the spin $z-$projection $\sigma=\pm 
1/2$. It is also known that actually we does not need to know the 
$N-$matrix in detail. In particular, to investigate the total system 
energy together with the kinetic and interaction ones, we can 
deal with the $2-$matrix defined by
{\setlength\arraycolsep{0pt}
\begin{eqnarray}
\varrho_2(x_1',x_2';x_1,x_2)\!=\!\int\limits_{V}\!&&d x_3\!\dots\!d 
x_N \Psi^{*}(x_1,x_2,x_3,\dots,x_N)\nonumber\\
&&\times\;\Psi(x_1',x_2',x_3, \dots ,x_N),
\label{2mA}
\end{eqnarray}}
where in general
$$
\int\limits_{V} \dots dx=\sum\limits_{\sigma}
\int\limits_{V} \dots d^3 r.
$$
Let us introduce the eigenfuctions of the $2-$matrix 
$\xi_{\nu}(x_1,x_2)$ given by
{\setlength\arraycolsep{0pt}
\begin{eqnarray}
\int\limits_{V}\!dx_1 dx_2\; 
\varrho_2(x_1',x_2';x_1,x_2)\xi_{\nu}&&(x_1,x_2)=\nonumber\\
=&& w_{\nu}\xi_{\nu}(x_1',x_2'),
\label{eigen}
\end{eqnarray}}where $w_{\nu}$ stands for the $\nu-$state eigenvalue. 
These eigenfuctions are usually called  in-medium PWF~\cite{bog}. 
The $2$-mat\-rix can be expressed in terms of its eigenfunctions and 
eigenvalues as follows:
\begin{equation}
\varrho_2(x_1,x_2;x_1',x_2')=\sum\limits_{\nu}w_{\nu}\;
\xi^{*}_{\nu}(x_1,x_2)\xi_{\nu}(x_1',x_2'),
\label{2mB}
\end{equation}
where it is implied that
$$
\int\limits_{V}\!dx_1 dx_2\;
\xi^{*}_{\nu}(x_1,x_2)\xi_{\nu'}(x_1,x_2)=\delta_{\nu\nu'}.
$$
From Eq.~(\ref{2mA}) it follows that
\begin{equation}
\int\limits_{V}\!dx_1 dx_2\; \varrho_2(x_1,x_2;x_1,x_2)=1,
\label{2mint}
\end{equation}
and, hence,
\begin{equation}
\sum_{\nu}w_{\nu}=1,
\label{norm1}
\end{equation} 
which allows one to interprete the eigenvalue $w_{\nu}$ as 
the probability of observing a particle pair in the $\nu-$state.

Now let us remind that the total momentum of the system of 
interest, the total system spin and its $z-$projection are 
conserved quantities~\cite{note2}. This means that they commute 
with the $N-$particle density matrix because the latter is 
permutable with the system Hamiltonian. As the total pair 
momentum $\hbar{\rm \hat{Q}}$, the total pair spin $\hat{S}$ and its 
${\rm Z}-$component $\hat{S}_Z$ commute with the total system 
momentum, total system spin and its ${\rm Z}-$projection, 
correspondingly, they commute with the $N-$matrix, too. If so, 
then one can derive that ${\rm \hat{Q}}$, $\hat{S}$ and $\hat{S}_Z$ 
are permutable with the $2-$matrix. This is why we can choose 
the eigenfunctions of the $2-$matrix in such a way 
that~\cite{bog} $\nu=\{\lambda,{\bf Q},S,m_S\}$, where 
$m_S$ is an eigenvalue of $S_Z$ and $\lambda$ stands for other 
quantum numbers. Hence, in the homogeneous situation one arrives 
at (see Refs.~\cite{bog} and \cite{shanC})
\begin{equation}
\xi_{\nu}(x_1,x_2)=\vartheta_{\nu}({\bf r},\sigma_1,\sigma_2)\,
\exp\left(i {\bf Q}{\bf R}\right)/\sqrt{V},
\label{pwfA}
\end{equation}
where ${\bf r}={\bf r}_1-{\bf r}_2$ and ${\bf R}=({\bf r}_1+{\bf 
r}_2)/2.$ As the in-medium bound pair states like the BCS-pairs are 
beyond the scope of the present publication, here we deal only 
with the scattering states. In other words, only the sector of the 
``dissociated'' pair states is taken into consideration. Hence, 
$\nu=\{{\bf q},{\bf Q},S,m_S\}$, where ${\bf q}$ stands for 
the relative wave vector. This is why it is convenient to set by 
definition
\begin{equation}
\vartheta_{\nu}({\bf r},\sigma_1,
\sigma_2)=\varphi_{\nu}({\bf r},\sigma_1,\sigma_2)/\sqrt{V}.
\label{pwfB}
\end{equation} 
The pair interaction of interest does not depend on the spin variables, 
which means that $\varphi_{\nu}({\bf r},\sigma_1,\sigma_2)$ is 
expressed as
\begin{equation}
\varphi_{\nu}({\bf r},\sigma_1,\sigma_2)=\varphi_{\;{\bf q},{\bf Q},S}
({\bf r})\,\chi_{S,m_S}\!(\sigma_1,\sigma_2),
\label{factor}
\end{equation}
where for the singlet states $(S=0)$ one gets
\begin{equation}
\chi_{0,0}\!(\sigma_1,\sigma_2)=\Delta(\sigma_1+\sigma_2) 
\mbox{sign}(\sigma_1)/\sqrt{2},
\label{singlet}
\end{equation}
while for the triplet wave functions $(S=1)$ 
{\setlength\arraycolsep{0pt}
\begin{eqnarray}
\chi_{1,m_S}&&(\sigma_1,\sigma_2)=\nonumber\\[1mm]
    &&=\left\{\begin{array}{ll}
\Theta(-\sigma_1) \Theta(-s_2)\quad  &{\rm if}\quad m_S=-1,\\[1mm]
\Delta(\sigma_1+\sigma_2)/\sqrt{2}\quad&{\rm if}\quad 
                                           m_S=\,\;\;0,\\[1.5mm]
\Theta(\sigma_1) \Theta(\sigma_2)\quad &{\rm if}\quad m_S=\,\;\;1.
\end{array}\right.
\label{triplet}
\end{eqnarray}}In Eqs.~(\ref{singlet}) and (\ref{triplet})
$$
\Delta(\sigma)=\left\{\begin{array}{ll}
0\;\;\;&{\rm if}\;\;\;\sigma\not=0, \\
1\;\;\; &{\rm if}\;\;\;\sigma=0,
\end{array}\right.\quad
\Theta(\sigma)
=\left\{\begin{array}{ll}
1 \;\;\;&{\rm if}\;\;\; \sigma \geq 0,\\
0 \;\;\;&{\rm if}\;\;\; \sigma < 0.
\end{array}\right.
$$
Now, from Eq.~(\ref{singlet}) it follows that
\begin{equation}
\chi_{0,0}(\sigma_1,\sigma_2)=-\chi_{0,0}(\sigma_2,\sigma_1).
\label{singletA}
\end{equation}
Then, the Fermi statistics dictates
\begin{equation}
\varphi_{{\bf q},{\bf Q},0}({\bf r})=
\varphi_{{\bf q},{\bf Q},0}({-\bf r})=
\varphi_{-{\bf q},{\bf Q},0}({\bf r}),
\label{singletB}
\end{equation}
and for $r \to \infty$ the wave function $\varphi_{{\bf 
q},{\bf Q},0}({\bf r})$ obeys the asymptotic regime
\begin{equation}
\varphi_{{\bf q},{\bf Q},0}({\bf r}) \to \sqrt{2} 
                                \cos({\bf q}{\bf r}).
\label{singletC}
\end{equation}
In turn, for triplet states Eq.~(\ref{triplet}) yields 
\begin{equation}
\chi_{1,m_S}(\sigma_1,\sigma_2)=\chi_{1,m_S}(\sigma_2,
\sigma_1),
\label{tripletA}
\end{equation}
which leads to 
\begin{equation}
\varphi_{{\bf q},{\bf Q},1}({\bf r})=
-\varphi_{{\bf q},{\bf Q},1}({-\bf r})=
-\varphi_{-{\bf q},{\bf Q},1}({\bf r}).
\label{tripletB}
\end{equation}
Now, another boundary regime 
\begin{equation}
\varphi_{{\bf q},{\bf Q},1}({\bf r}) \to \sqrt{2} 
                                 \sin({\bf q}{\bf r}).
\label{tripletC}
\end{equation}
is fulfilled when $r \to \infty$.

Working in the thermodynamic limit $N\to \infty,V\to \infty,
N/V=n=const$, it is more convenient to leave the $2-$matrix 
in favour of the so-called  pair correlation function 
\begin{equation}
F_2(x_1,x_2;x_1',x_2')=\bigl\langle\hat{\psi}^{\dagger}(x_1)
\hat{\psi}^{\dagger}(x_2)\hat{\psi}(x_2')\hat{\psi}(x_1')\bigr
\rangle,
\label{2mC}
\end{equation} 
where $\langle\hat{A}\rangle$ stands for the statistical average of 
the operator $\hat{A}$, and $\hat{\psi}^{\dagger}(x)$, $\hat{\psi}(x)$ 
denote the field Fermi operators. The pair correlation function 
differs from the $2-$matrix by the normalization factor~(see 
Ref.~\cite{bog}),
\begin{equation}
F_2(x_1,x_2;x_1',x_2')=N(N-1)\varrho_2(x_1',x_2';x_1,x_2),
\label{2mD}
\end{equation}
so that $F_2$ remains finite while $\varrho_2$ approaches zero in the 
thermodynamic limit. Indeed, when $V \to \infty,\,N \to \infty,\,
N/V=n\to const$, Eqs.~(\ref{2mB}) and (\ref{2mD}), taken in 
conjunction with Eqs.~(\ref{pwfA}) and (\ref{pwfB}), yield
{\setlength\arraycolsep{0pt}
\begin{eqnarray}
&&F_2(x_1,x_2;x_1',x_2')=\nonumber\\
&&=\!\!\sum\limits_{S,m_S}\!\int \frac{d^3\!{\rm q}\,d^3\!{\rm 
Q}}{(2\pi)^6}\;\rho_{S,m_S}\!({\bf q},{\bf Q})
\varphi^{*}_{{\bf q},{\bf Q},S}({\bf r})\varphi_{{\bf q},{\bf 
Q},S}({\bf r}')\nonumber\\&&\times\;\chi^{*}_{S,m_S}\!(\sigma_1,
\sigma_2)\chi_{S,m_S}\!(\sigma_1',\sigma_2')\exp\left\{i{\bf Q}
({\bf R}'\!-\!{\bf R})\right\},
\label{2mE}
\end{eqnarray}}
where the momentum-distribution function
\begin{equation}
\rho_{S,m_S}\!({\bf q},{\bf Q})=\lim_{\;V,N\to \infty}
\Bigl\{N(N-1) w_{{\bf q},{\bf Q},S,m_S}\Bigr\}
\label{rhosms}
\end{equation}
is finite because $w_{{\bf q},{\bf Q},S,m_S} \sim 1/V^2$~(this follows
from Eq.~(\ref{norm1}) when $V \to \infty$). For $\rho_{S,m_S}\!
({\bf q},{\bf Q})$ one gets the relation   
\begin{equation}
\sum\limits_{S,m_S}\int\frac{d^3\!{\rm q} d^3\!{\rm Q}}{(2\pi)^6} 
\;\rho_{S,m_S}\!({\bf q},{\bf Q})=n^2
\label{norm2}
\end{equation}
resulting from Eqs.~(\ref{norm1}) and (\ref{rhosms}). 

All the necessary formulae are now discussed and displayed, and one 
can turn to calculations of the interaction energy. Using the 
well-known expression
\begin{equation}
E_{\rm int}=\frac{1}{2} \int dx_1 dx_2 V(|{\bf r}_1-
{\bf r}_2|)F_2(x_1,x_2;x_1,x_2)
\label{eint3}
\end{equation}  
and keeping in mind Eq.~(\ref{2mE}), one gets the following 
important relation:
{\setlength\arraycolsep{0pt}
\begin{eqnarray}
\varepsilon_{\rm int}&=&\frac{1}{2n}\!\int\! d^3r\; V(r)
\nonumber\\&\times&\;\sum\limits_{S,m_S}\!\int\frac{d^3\!
{\rm q}\,d^3\!{\rm Q}}{(2\pi)^6}\; \rho_{S,m_S}\!({\bf q},
{\bf Q})\;|\varphi_{{\bf q},{\bf Q},S}({\bf r})|^2,
\label{eint4}
\end{eqnarray}}
provided the equality
$$
\sum_{\sigma_1,\sigma_2}\chi^{*}_{S,m_S}\!(\sigma_1,\sigma_2)
\chi_{S,m_S}\!(\sigma_1,\sigma_2)=1
$$  
is taken into account. Equation (\ref{eint4}) directly connects the 
interaction energy per fermion with PWF and, thus, with the 
scattering waves defined by
\begin{equation}
\psi_{{\bf q},{\bf Q},0}({\bf r})=\varphi_{{\bf q},{\bf Q},0}
({\bf r})-\sqrt{2}\cos({\bf q}{\bf r})
\label{singletD}
\end{equation}
and
\begin{equation}
\psi_{{\bf q},{\bf Q},1}({\bf r})=\varphi_{{\bf q},{\bf Q},1}
({\bf r})-\sqrt{2}\sin({\bf q}{\bf r}).
\label{tripletD}
\end{equation}
The scattering waves are immediately related to the 
pairwise-potential contribution to the spatial particle 
correlations. Setting $\psi_{{\bf q},{\bf Q},0}({\bf r})=\psi_{{\bf 
q},{\bf Q},1}({\bf r})=0$, or, in other words, ignoring that 
contribution and taking notice only of the correlations due to 
the statistics, one arrives at the Hartree-Fock scheme.

So far we did not invoke any approximation when ope\-rating with
the $2-$matrix and pair correlation function~\cite{note3}. However, 
taken in the regime of a dilute Fermi gas, Eq.~(\ref{eint4}) can 
significantly be simplified. Indeed, the lower densities, the lower 
momenta are typical of the system. This means that the pair momentum
distibution $\rho_{S,m_S}({\bf q},{\bf Q})$ is getting more localized 
in a small vicinity of the point ${\rm q}={\rm q}=0$ when $n \to 
0$. Consequently, the low-momentum approximation can be applied 
according to which for $n \to 0$ we get 
{\setlength\arraycolsep{0pt}
\begin{eqnarray}
\int\!\frac{d^3\!{\rm q}\,d^3\!{\rm Q}}{(2\pi)^6}\;
\rho_{S,m_S}\!({\bf q},{\bf Q})\;
|\varphi_{{\bf q},{\bf Q},S}&&({\bf r})|^2\!
\simeq\nonumber\\
\simeq &&|\varphi_{S}({\bf r})|^2 \; \eta_{\,S,m_S},
\label{lowp}
\end{eqnarray}}
where
\begin{equation}
\eta_{\,S,m_S}\!=\!\!\int\!\frac{d^3\!{\rm q}\,d^3\!{\rm Q}}{(2
\pi)^6}\;\rho_{S,m_S}({\bf q},{\bf Q})
\label{etaA}
\end{equation}
and
\begin{equation}
\varphi_{S}({\bf r})=\lim\limits_{{\rm q,Q}\to 0}
\varphi_{{\bf q},{\bf Q},S}({\bf r}).
\label{vars}
\end{equation}
From Eqs.~(\ref{tripletB}) and (\ref{vars}) it follows that 
$\varphi_{S=1}({\bf r})=0$. This result, taken in conjunction 
with the low-momentum approximation of Eq.~(\ref{lowp}), makes 
it possible to conclude that Eq.~(\ref{eint4}) reduces for 
$n \to 0$ to 
\begin{equation}
\varepsilon_{\rm int}\simeq\frac{\eta_{\,0,0}}{2n}
\!\int\!d^3r V(r)|\varphi_{0}({\bf r})|^2.
\label{eint5}
\end{equation}
The triplet states do not make any contribution to the 
interaction energy in the approximation (\ref{lowp}), and 
this completely meets the usual expectations. 

Now, to employ Eq.~(\ref{eint5}), one should have an idea 
concerning $\varphi_{0}({\bf r})$ and $\eta_{\,0,0}$. As to the 
limiting wave function $\varphi_{0}({\bf r})$, it can be 
determined by means of the following simple and custom arguments. 
In the dilution limit the pair wave function $\varphi_{{\bf q},
{\bf Q},S}({\bf r})$ approaches the solution of the ordinary 
two-body Schr\"odinger equation
{\setlength\arraycolsep{0pt}
\begin{eqnarray}
-(\hbar^2/m)\nabla^2\varphi_{{\bf q},{\bf Q},S}({\bf r})&&+
V(r)\varphi_{{\bf q},{\bf Q},S}({\bf r})=\nonumber\\[1mm]
&&=(\hbar^2 q^2/m)\varphi_{{\bf q},{\bf Q},S}({\bf r})
\label{schr1}
\end{eqnarray}}with the boundary conditions given by 
Eqs.~(\ref{singletB}) and (\ref{tripletB}). Hence, in the limit 
$n \to 0$ the quantity $\varphi_0(r)$ has to obey the equation 
\begin{equation}
-(\hbar^2/m)\nabla^2\varphi_{0}({\bf r})+V(r)\varphi_{0}({\bf r})=0,
\label{schr2}
\end{equation}
where $\varphi_{0}({\bf r})\to \sqrt{2}$ when $r \to \infty$. Comparing 
Eq.~(\ref{schr}) with Eq.~(\ref{schr2}), for $n \to 0$ one finds 
\begin{equation}
\varphi_{0}({\bf r})=\sqrt{2}\,\varphi(r).
\label{var0}
\end{equation}   
To complete calculation of the interaction energy, it only 
remains to find $\eta_{\,0,0}$. One can expect that when numbers 
of fermions with positive and negative spin $z-$projections are the 
same, the magnitude of $\rho_{S,m_S}({\bf q},{\bf Q})$ appears to be 
independent of the spin variables. In this case Eqs.~(\ref{norm2})
and (\ref{etaA}) give
\begin{equation}
\eta_{S,m_S}=n^2/4.
\label{etaB}
\end{equation}
Note that Eq.~(\ref{etaB}) can readily be found in a more rigorous 
way concerning the relation
\begin{equation}
\frac{1}{V^2}\int d^3r_1\,d^3r_2 F_2(x_1,x_2;x_1,x_2)=
n_{\sigma_1}n_{\sigma_2},
\label{etaC}
\end{equation}
where $n_{\sigma}=\langle\hat{\psi}^{\dagger}(x_1)\hat{\psi}(x_1)
\rangle$. This relation results from the definition of the pair 
correlation function (\ref{2mC}). To derive Eq.~(\ref{etaB}) from
Eq.~(\ref{etaC}), one should employ the latter in conjunction with
Eqs.~(\ref{singlet}), (\ref{triplet}) and (\ref{2mE}) and, then, take 
account of $n_{\sigma}=\langle\hat{\psi}^{\dagger}(x_1)\hat{\psi}(x_1)
\rangle=n/2$. Let us stress that Eq.~(\ref{etaB}) is not general. 
For example, when all the considered fermions have the spin 
$z-$projection equal to $+1/2$, one gets $\eta_{1,1}=n^2$ and 
$\eta_{0,0}=\eta_{1,-1}=\eta_{1,0}=0.$ As it is seen, in this case 
the interaction energy resulting from Eq.~(\ref{eint5}) is exactly 
amount to zero: one should go beyond the approximation defined by 
Eq.~(\ref{lowp}) to get an idea about $\varepsilon_{\rm int}$ of 
such a weekly interacting system. Here it is worth remarking that 
this week interaction is an obstacle that can prevent experimentalists 
from observing possible BCS-like pairing of fermions due to an extremely
low temperature of the BCS-transition. To strengthen the interaction 
effects, it was, in particular, suggested~\cite{stoof} to complicate 
the experimental scheme in such a way that fermions with various spin 
$z-$projections would be trapped. In this case the low-momentum 
approximation yields a finite result for $\varepsilon_{\rm int}$. 
It is deductive to go in more detail concerning this situation 
because here another choice of the eigenfunctions of the 
$2-$matrix turned out to be convenient rather than that of 
Eq.~(\ref{factor}). The details are in Appendix.

At last, inserting Eqs.~(\ref{var0}) and (\ref{etaB}) in 
Eq.~(\ref{eint5}) and making use of Eqs.~(\ref{aB}) and (\ref{bA}), 
in the dilution limit one can derive $\varepsilon_{\rm int}\simeq\pi
\hbar^2 n (a-b)/m$, which is nothing else but the leading term in 
Eq.~(\ref{eint2}). Note that to derive the next-to-leading terms in
the expression for the interaction energy via PWF, one should 
construct a more elaborated model similar to that of 
Refs.~\cite{shanA,shanB} concerning a dilute Bose gas. The model like 
that has to take into account in-medium corrections to PWF
to go beyond the approximation of Eq.~(\ref{lowp}). Though this 
investigation is beyond the scope of the present publication, there 
are some important remarks on the in-medium corrections to the 
eigenfunctions of the $2-$matrix in the next section.

Here it is of interest to check if Eq.~(\ref{eint5}) yields 
Eq.~(\ref{eintPS}) when replacing $V(r)$ by the pseudopotential 
$V^{(\rm ps)}(r)$. The simplest way of escaping divergences while 
operating with the pseudopotential is to adopt $V^{(\rm ps)}(r)=
(4\pi\hbar^2a/m)\delta({\bf r})$ in conjunction with the 
Hartree-Fock scheme. For example, exactly this way was used in the 
classical paper by Pitaevskii when deriving the well-known 
Gross-Pitaevskii equation for the order parameter of the 
Bose-Einstein condensation in a dilute Bose gas~\cite{gp}. A more 
elaborated variant, which goes beyond the Hartree-Fock framework, 
requires a more sophisticated choice of the pseudopotential which, 
for the particular case of the hard-sphere interaction, is of the 
form $V^{(\rm ps)}(r)=(4\pi\hbar^2a/m)\delta({\bf r})
(\partial/\partial r)r$~\cite{hy}. The aim of this variant 
is to calculate not only the total system energy but some 
additional important characteristics (for instance, the pair 
correlation function) which can not be properly considered in
the former way. Complicating the pseudopotential construction 
allows one to escape a double account of some scattering channels 
(this fact is known since the Thesis by Nozi\'eres). This double 
account appears due to the fact that the particle scattering 
makes contribution to the pseudopotential. In our case, of course, 
the simplest choice is enough. Now, replacing $V(r)$ 
by $V^{(\rm ps)}(r)=(4\pi\hbar^2a/m) \delta({\bf r})$ in 
Eq.~(\ref{eint5}) and setting $\psi(r)=0\;(\varphi(r)=1)$, for 
$n \to 0$ one derives $\varepsilon^{(\rm ps)}_{\rm int}\simeq 
(\pi\hbar^2 a\,n/m)$. It is just the leading term in 
Eq.~(\ref{eintPS}). This supports the conclusion that the 
pseudopotential is not able to produce correct results for the 
kinetic and interaction energies of dilute quantum gases. 

Here let us make some remarks on the momentum distribution 
of the ``dissociated'' pairs $\rho_{S,m_S}({\bf q},{\bf Q})$. 
The calculational procedure leading to Eq.~(\ref{eint5}) 
does not involve a detailed knowledge of this distribution. 
However, it can be completely refined. In Ref.~\cite{shanC} it 
was suggested to derive $\rho_{S,m_S}({\bf q},{\bf Q})$ via the 
correlation-weakening principle (CWP). According to CWP the pair 
correlation function obeys the following relation: 
\begin{eqnarray}
F_2(x_1,x_2;x_1',x_2') \to F_1(x_1;x_1')\,F_1(x_2;x_2')
\label{cwpA}
\end{eqnarray}
when
$$
|{\bf r}_1-{\bf r}_2|\to\infty,\;
|{\bf r}_1-{\bf r}_1'|=const,\;|{\bf r}_2-{\bf r}_2'|=const.
$$
In Eq.~(\ref{cwpA}) we set $F_1(x_1;x_1')=\langle\hat{\psi}^{\dagger}
(x_1)\hat{\psi}(x_1')\rangle$. So, the pair
momentum distribution $\rho_{S,m_S}({\bf q},{\bf Q})$, which 
appears in the expansion of $F_2$ in the set of its eigenfunctions, 
can be expressed in terms of the single-particle momentum distribution 
$n_{\sigma}(k)=\langle a^{\dagger}_{\sigma}({\bf k})a_{\sigma}
({\bf k})\rangle$, that comes into the plane-wave expansion for 
$F_1$. In the case of interest, when the both distribution functions 
turn out to be independent of spin variables, this leads to
\begin{equation}
\rho_{S,m_S}({\bf q},{\bf Q})=n\bigl(|{\bf Q}/2+{\bf q}|\bigr)\,
n\bigl(|{\bf Q}/2-{\bf q}|\bigr),
\label{cwpE}
\end{equation} 
where, by definition, $n(k)=n_{\sigma}(k)$. Concluding let us 
set, for the sake of demonstration, $n(k)=1$ for $k\leq k_F$, 
$n(k)=0$ for $k > k_F$ and return to Eq.~(\ref{lowp}). Inserting 
Eq.~(\ref{cwpE}) in the right-hand side of Eq.~(\ref{lowp}) and 
utilizing this single-particle momentum distribution of an ideal 
Fermi gas, we arrive at the left-hand side of Eq.~(\ref{lowp}) due 
to $k_F \to 0$ when $n \to 0$. This example is a good illustration 
of the idea of the low-momentum approximation introduced by 
Eq.~(\ref{lowp}).

Thus, in Section \ref{sec4} it is demonstrated how to calculate the
interaction energy of a dilute Fermi gas from the first principles,
beyond the formula by Huang and Yang taken in conjunction with the 
Hellmann-Feynman theorem. Though the results of this section make
Eq.~(\ref{eint2}) physically sound and support the conclusion about
strong influence of the pairwise interaction on the system kinetic 
energy, the nature of this influence is not highlighted yet. The 
detailed discussion of this nature is given in Section~\ref{sec5}. 

\section{Kinetic energy via the pair wave functions}
\label{sec5}

Some hints as to how to proceed with the problem of the influnce of
the pairwise particle interaction on the kinetic energy can be found 
in the Bogoliubov model of a weakly interacting Bose gas and in the 
BCS-approach. As it shown in Refs.~\cite{shanA,shanB}, there exists 
some important relation which mediates between the pairwise boson 
interaction and single-boson momentum distribution $n_B(k)$ in the 
Bogoliubov model. For the ground-state case this relation is the 
form
\begin{equation}
n_{B}(k)\bigl[1+n_{B}(k)\bigr]=n^2_0 \psi_{B}^2(k),
\label{bog}
\end{equation}  
where $\psi_B(k)$ is the Fourier transform of the scattering part 
of the bosonic PWF corresponding to ${\bf q}={\bf Q}=0$, and $n_0$
stands for the density of condensed bosons. When the pairwise boson
interaction is ``switched off'', there is no scattering. So, bosonic 
PWF are the symmetrized plane waves and $\psi_B(k)=0$. In this case 
Eq.~(\ref{bog}) has the only physical solution $n_B(k)=0$, 
that corresponds to an ideal Bose gas with the zero condensate 
depletion and the zero kinetic energy. On the contrary, ``switching 
on'' the pairwise interaction leads to $\psi_B(k)\not=0$, and we 
arrive at the regime of a nonzero condensate depletion, when 
$n_B(k)\not=0$ and the kinetic energy is not equal to zero, 
as well. 

A similar situation occurs in the BCS-model. There is again some 
corner-stone relation mediating between the pairwise interaction 
and the single-particle momentum distribution $n_{BCS}(k)$. It can 
be expressed\cite{shanC} as
\begin{equation}
n_{BCS}(k)\bigl[1-n_{BCS}(k)\bigr]=n^2_s\,\psi_{BCS}^2(k),
\label{bcs}
\end{equation}      
where $\psi_{BCS}(k)$ is the Fourier transform of the internal wave
function of a condensed bound pair of fermions, $n_s$ is the density 
of this pairs.``Switching off'' the pairwise attraction leads to 
disappearnce of the bound pair states: $\psi_{BCS}(k)=0$. In this case 
there are two branches of the solution of Eq.~(\ref{bog}): $n_{BCS}=1$ 
and $n_{BCS}=0$. Below the Fermi momentum the first branch is advantageous 
from the thermodynamic point of view, while the second one is of 
relevance above. So, one gets the regime of an ideal Fermi gas with the 
familiar kinetic energy, often called the Fermi energy. When ``switching 
on'' the attraction, some significant corrections to the momentum 
distribution of an ideal Fermi gas arise. This corrections are 
dependent of the mutual attraction of fermions and make a significant 
contribution to the kinetic energy additional to the Fermi energy. 

Now, keeping in mind the examples listed above, one can suppose that 
the relation connecting PWF (strictly speaking, the scattering waves 
and bound waves) with the single-particle momentum distribution is 
some general feature of quantum many-body systems. If so, exactly 
this relation has to be responsible for the influence of the pairwise 
interaction on the kinetic energy of the quantum gases. For a 
ground-state dilute Fermi gas with no pairing effects the relation of 
interest can be constructed in the form  
\begin{equation}
n(k)[1-n(k)]={\cal L}(k),
\label{fermi_gas}
\end{equation}  
where ${\cal L}(k)$ stands for a functional of the in-medium scattering 
waves $\psi_{{\bf q},{\bf Q},S}({\bf r})$. To go in more detail 
concerning the functional ${\cal L}(k)$, we should try to calculate the 
kinetic energy, starting durectly from $n(k)$ of Eq.~(\ref{fermi_gas}). 
This equation results in
\begin{equation}
n(k)=\left\{\begin{array}{ll}
\bigl[1+\sqrt{1-4{\cal L}(k)}\bigr]/2 &\;\;{\rm if}  \quad
k \leq {\cal K}_F,\\[1.5mm]
\bigl[1-\sqrt{1-4{\cal L}(k)}\bigr]/2\; &\;\;{\rm if}\quad
k > {\cal K}_F,
\end{array}\right.
\label{nkA} 
\end{equation}
where $\hbar{\cal K}_F$ stands for the Fermi momentum. In the present 
paper a weakly nonideal gas of fermions is under 
investigation~\cite{note4}, which means that the single-fermion 
momentum distribution approaches the ideal-gas Fermi distribution 
in the dilution limit: ${\cal L}(k)\to 0$ and ${\cal K}_F\to k_F$ 
for $n \to 0$. Then, Eq.~(\ref{nkA}) can be rewritten for $n \to 0$ 
as 
\begin{equation}
n(k)\simeq \bigl[1-\ell(k)\bigr]\Theta(k_F-k)+\ell(k)\Theta(k-k_F),
\label{nkB}
\end{equation}
where the dilution expansions ${\cal L}(k)=\ell(k)(1+\ldots)$ and
${\cal K}_F=k_F(1+\ldots)$ are implied. Taken together with the 
familiar formula
\begin{equation}
E_{\rm kin}=V\sum_{\sigma}\int\frac{d^3k}{(2\pi)^3}\;T_k n_{\sigma}(k),
\label{ekin3}
\end{equation}
Eq.~(\ref{nkB}) leads for $n \to 0$ to  
\begin{equation}
\varepsilon_{\rm kin}=\frac{3\hbar^2 k^2_F}{10m}+\frac{2}{n}
\int\frac{d^3k}{(2\pi)^3} T_k \,\ell(k)+\ldots.
\label{ekin4}
\end{equation}
Note that the characteristic length $b$ given by (\ref{bA}) can be 
rewritten as
$$
b=\frac{m}{2\pi \hbar^2}\int\frac{d^3k}{(2\pi)^3} T_k \psi^2(k),
$$
where $\psi(k)$ is the Fourier transform of the scattering wave 
$\psi(r)$~(see Eq.~(\ref{as})). Keeping this in mind and comparing 
Eqs.~(\ref{ekin2}) with (\ref{ekin4}), we can find for $n \to 0$ 
that
\begin{equation}
\ell(k)=(n^2/4)\psi^2(k).
\label{LA}
\end{equation} 
Hence, the quantuty ${\cal L}(k)$ is indeed a functional
of $\psi_{{\bf p},{\bf q},S}({\bf r})$ that reduces to the 
right-hand side of Eq.~(\ref{LA}) in the limit $n \to 0$.
So, our expectations about the relation mediating between the 
pairwise interaction and single-particle momentum distribution 
in a dilute Fermi gas turn out to be adequate. Let us remark 
that Eq.~(\ref{nkB}) is a good approximation only when calculating 
the dilution expansion for the kinetic energy (strictly speaking, 
the leading and next-to-leading terms). However, to go in more
detail concerning the single-fermion momentum distribution, one 
should be based on Eq.~(\ref{nkA}) rather than on Eq.~(\ref{nkB}). 
Indeed, Eq.~(\ref{schr}) can be rewritten as $$\psi(k)=-\frac{1}{2T_k}
\int d^3r \varphi(r) V(r)\exp(-i{\bf k}{\bf r}).$$ From Eq.~(\ref{aA}) 
it follows that $\int d^3r \varphi(r) V(r)=4\pi\hbar^2a/m$. Therefore, 
$\psi(k)\propto -1/k^2$ when $k \to 0$. Taken in the first two orders 
of the dilution expansion, the kinetic energy is not affected by this 
singularity. However, $n(k)$ is rather sensitive to it. Actually 
in-medium corrections to $\psi(k)$ should be involved to find the 
adequate behaviour of $n(k)$ when $k\to \infty$. Indeed, low momenta 
correspond to large particle separations where influence of the 
surrounding medium on PWF can be of importance even in the dilution 
regime.

Discussion about the in-medium corrections to PWF can be illustrated 
as follows. Keeping in mind the replacement of $\varphi_0(r)$ by 
$\sqrt{2}\varphi(r)$ in Eq.~(\ref{eint5}), one can suggest to abandon 
the approximation ${\cal L}(k)\simeq (n^2/4)\psi^2(k)$ in favour of 
\begin{equation}
{\cal L}(k)\simeq (n^2/8)|\psi_0(k)|^2,
\label{LB}
\end{equation}
where $\psi_0(k)$ is the Fourier transform of $\psi_0(r)=\varphi_0(r)
-\sqrt{2}$. As $\psi_0(k) \to \sqrt{2}\psi(k)$ for $n \to 0$, the 
correct dilution limit ${\cal L}(k) \to \ell(k)$ results from 
Eq.~(\ref{LB}). However, the advantage of Eq.~(\ref{LB}) as compared 
with ${\cal L}(k)\simeq \ell(k)$ is that $\psi_0(k)$ is an in-medium 
PWF and, so, includes in-medium corrections at any finite particle 
density. This is why one can expect that Eq.~(\ref{LB}), taken 
together with Eq.~(\ref{nkA}), gives adequate results for both the 
kinetic energy and the single-fermion momentum distribution. To be
convinced of this, let us try first to have a guess about the in-medium 
corrections to PWF. Here it is natural to start from the expectation 
that the Fermi sphere is completely occupied when $n \to 0$. This 
means (see Eqs.~(\ref{fermi_gas}) and (\ref{LB})) that $\psi_0(k)=0$ 
at $k\leq {\cal K}_F\simeq k_F$. On the contrary, at large $k$ (small 
particle separations!) $\psi_0(k)$ can not be significantly affected 
by surroundings. So, $\psi_0(k)$ in this regime is nearly be governed 
by the ordinary two-body Schr\"odinger equation for the internal wave 
function of the interacting pair at ${\bf q}=0$. Combining these two 
regimes, one can approximately write
\begin{equation}
2T_k\psi_0(k)+\Theta(k\!-\!k_F)\!\!\!\int\!\! d^3r
\varphi_0(r)\!V(r)\!\exp(-i{\bf k}{\bf r})=0.
\label{BBGa}
\end{equation}   
Stress that the jump in the Fourier transform of $\psi_0(r)$ at 
$k=k_F$ does not imply a jump of $\psi_0(r)$ itself. Passing 
to the coordinate representation, Eq.~(\ref{BBGa}) gets the form 
{\setlength\arraycolsep{0pt}
\begin{eqnarray}
-\frac{\hbar^2}{m}\nabla^2\varphi_0(r)+&&\varphi_0(r)V(r)=\nonumber\\
=&&\int d^3r'\varphi_0(r')V(r') G({\bf r}-{\bf r}'),
\label{BBGb}
\end{eqnarray}}where $G({\bf r})=\int_{k<k_F} d^3k/(2\pi)^3
\exp[i{\bf k}{\bf r}]$. Equation~(\ref{BBGb}) is nothing else but the 
simplest version of the Bethe-Goldstone equation~\cite{bg} for the 
internal wave function of the interacting pair at ${\bf Q}={\bf q}
=0$. So, the ``veto'' upon appearance of intermediate scattering 
states inside the Fermi sphere~(see the page {\it 320} in the 
textbook~\cite{fw}), that allows for passing from the ordinary 
two-body Schr\"odinger equation to the Bethe-Goldstone one, can be 
easily explained beyond any intuitive arguments with the help of the 
relation connecting the scattering parts of the in-medium PWF with 
the single-particle momentum distribution.  

Now, after the glance at the problem of medium influence on PWF, 
we can try to outline a more elaborated way of treating the 
in-medium pair wave functions. Indeed, it turns out that there are
everything at our disposal to derive some two coupled equations 
which make it possible to find $\psi_0(k)$ in conjunction with 
$n(k)$. The first of these equations connecting $\psi_0(k)$ with
$n(k)$ results from Eqs.~(\ref{nkA}) and (\ref{LB}). As to the 
second equation, some longer but straightforward calculations are
needed. The point is that Eqs.~(\ref{eint5}) and (\ref{ekin3}) 
enable representation of the total system energy as a functional 
of $\varphi_0(r)$ and $n(k)$. Hence, we are able to derive the 
second equation from the minimum condition for this functional.
Making variation with respect to $\psi^*_0(r)$ and $n(k)$, one 
gets
{\setlength\arraycolsep{0pt}
\begin{eqnarray}
\frac{\delta E}{V}=\int\frac{d^3k}{(2\pi)^3} &&\Bigl[2T_k \delta n(k)
+\frac{n^2}{8}\delta\psi^*_0(k)\nonumber\\
&&\times\int d^3r \varphi_0(r)V(r)\exp(-i{\bf k}{\bf r})\Bigr].
\label{dE}
\end{eqnarray}}To extract the second equation from Eq.~(\ref{dE}), 
one should keep in mind that the infinitesimal changes $\delta n(k)$ 
and $\delta\psi^*_0(k)$ are not independent due to the first equation. 
They are related to one another by
\begin{equation}
[1-2n(k)]\delta n(k)=(n^2/8)\psi_0(k) \delta \psi^*_0(k).
\label{dndpsi}
\end{equation} 
In addition, a change of $n(k)$ should not affect the total number of 
fermions $N=2\int d^3k/(2\pi)^3 n(k)$. So, the equation of interest 
has the form
\begin{equation}
\delta (E-\mu N)=0,
\label{GA}
\end{equation}  
where $\mu$ stands for the chemical potential. A combination of 
Eqs.~(\ref{dE}), (\ref{dndpsi}) and (\ref{GA}) yields
\begin{equation}
\psi_0(k)=\frac{1-2n(k)}{2(\mu-T_k)}\int d^3r \varphi_0(r) V(r) 
\exp(-i{\bf k}{\bf r}).
\label{GB}
\end{equation}
Equation~(\ref{GB}) is more complicated but very similar to 
Eq.~(\ref{BBGa}). This is especially clear in the light of 
the fact that for $n \to 0$ one gets $n(k) \to \Theta(k_F-k)$. 
As it is seen, when $k \to \infty$, Eq.~(\ref{GB}) is reduced 
to the ordinary Schr\"odinger equation for the internal wave 
function of an interacting pair with ${\bf q}=0$. While in the 
opposite regime, for $k\to 0$, Eq.~(\ref{GB}) acquires completely 
different form with significant contribution of medium-dependent 
terms. So, in-medium corrections are indeed of importance for getting 
an adequate behaviour of $n(k)$ and $\psi_0(k)$ at small absolute 
values of the wave vector ${\bf k}$. It is worth noting that the 
equation derived by Galitskii~\cite{gal} for the scattering part of his 
``effective wave function of two particles in medium''(see Eq.~(11.39) 
in Ref.~\cite{fw}) exactly reduces to Eq.~(\ref{GB}) in the case when 
the center-of-mass and relative wave vectors are equal to zero. The 
author of Ref.~\cite{gal} introduced the name ``effective wave 
function'' only due to similarity of the equation for that quantity to 
the Schr\"odinger equation for two particles in free space. To the best 
knowledge of the present authors, Galitskii did not associate those 
effective wave functions with the eigenfunctions of the $2-$matrix. 
However, the derived result strongly suggests that they are actually 
the eigenfunctions of the reduced density matrix of the second order. 
If so, there is a promising possibility of using the Galitskii 
equation in combination with the $2-$matrix, which can produce 
correct results for $\varepsilon$ and aslo for $\varepsilon_{\rm 
kin}$ and $\varepsilon_{\rm int}$. Note that in the case of a 
dilute Bose gas the pseudopotential approach does not yield correct 
picture of the short-range boson correlations in addition to the 
failure with the kinetic and interaction energies~\cite{shanA}. 
This is why one can expect the same fault with the pseudopotential 
in the Fermi case. In view of this fact, it is worth displaying one 
more advantage of the formalism of PWF. It produces the correct picture 
of the short-range spatial correlations. Indeed, from Eq.~(\ref{2mE}) 
it follows that the pair distribution function $g(r)$, defined by
$g(r)=(1/n^2)\sum_{\sigma_1,\sigma_2}\!\!F_2(x_1,x_2;x_1,x_2)$, can 
be expressed as
\begin{equation}
g(r)=\frac{1}{n^2}\!\!\int\!\frac{d^3\!{\rm q}\,d^3\!{\rm Q}}{(2\pi)^6}\;
\rho_{S,m_S}\!({\bf q},{\bf Q})\;|\varphi_{{\bf q},{\bf Q},S}
({\bf r})|^2.
\label{grA}
\end{equation}
When using the low-momentum approximation together with Eqs.~(\ref{var0})
and (\ref{etaB}), one gets that for $r \ll 1/k_F$ and $n \to 0$ 
Eq.~(\ref{grA}) reads
\begin{equation}
g(r)\simeq(1/2)\varphi^2(r).
\label{grB}
\end{equation}
This result differs from the pair distribution function of a dilute 
Bose gas~\cite{shanA,shanB} by the factor $1/2$ appearing precisely
due to the Fermi statistics. Note that this factor manifests itself 
in the total system energy, too. Indeed, it is well-known that the 
leading term in the dilution expansion for the total energy of a 
ground-state uniform Bose gas is twice more than the first 
$a-$dependent term in the corresponding expansion for a Fermi gas. 
Thus, contrary to the pseudopotential calculations for a Bose 
gas~(see Ref.~\cite{shanA}), there is no negative values of $g(r)$ 
at small particle separations when using the formalism of the 
$2-$matrix. At last, we remark that it would be also of importance 
to work out some arguments which would make it possible to refine 
the form of ${\cal L}(k)$, say, the ``imperfection'' functional. This 
is of interest in view of moving beyond the dilution regime in the 
investigations of the quantum many-body systems.
   
\section{Conclusion and comments}
\label{sec6}

Concluding, let us highlight the most important points of this 
article. 

First, the low-density expansions for the kinetic and interaction 
energies of a uniform ground-state Fermi gas with a repulsive 
pairwise interaction have been calculated up to the third order in 
the dilution parameter $k_Fa$. These quantities turn out to depend 
on the interparticle potential trough the two characteristic lengths 
$a$ and $b$ given by Eqs.~(\ref{aA}) and (\ref{bA}). In the first 
orders the $b-$dependent terms are cancelled in the sum of 
$E_{\rm kin}+E_{\rm int}$, and, so, the total energy $E$, if taken 
in those orders, involves only the scattering length $a$, as it is 
well-known. However, all the higher orders of the dilution expansion 
for $E$ are expected to emonstrate $b-$dependence. 

Second, the calculations have been fulfilled in two ways with the aim 
of controlling our conclusions and making the present investigation 
more physically sound. One of those ways is based on the 
Hellmann-Feynman theorem which is utilized in conjunction with the 
useful variational theorem for the scattering length when deriving 
$E_{\rm kin}$ and $E_{\rm int}$ from the familiar result by Huang 
and Yang for the total energy of a ground-state dilute uniform Fermi 
gas. Another way invokes a formalism of the in-medium pair wave 
functions. This variant allows one to find $E_{\rm kin}$ and
$E_{\rm int}$ starting directly from the first principles. The
advantage of the latter way is that it makes the underline physics 
more clear and enables to go in more detail concerning the spatial 
fermion correlations and their influence on the quantities of interest.

Third, the ratio of the two characteristic lengths $b/a$ has been 
investigated for the model pairwise interaction often used in the 
context of the alkali-metal atoms. In particular, according to the 
found results, one faces {\it a rather interacting gas of fermions} 
in the situation when $a=0$. The matter is that the sum of the terms 
coming from the pairwise interaction is here exactly cancelled in the 
total energy but, nevertheless, they make significant contributions 
to $E_{\rm kin}$ and $E_{\rm int}$, taken separately. This is a 
consequence of presence of the two characteristic lengths: when $a$ is 
amount to zero, then $b$ can take any positive value depending on a 
particular form of a pairwise interaction. Moreover, $b$ can be 
extremely large in such an ``uninteracting'' gas. Thus, one should be 
very careful with an experimantal analysis of a spatial expansion of a 
trapped gases with $a=0$~(see Ref. \cite{bourdel,ohara}). Actually in 
such situations the interaction energy can be extremely large in 
absolute value and has the negative sign.

Forth, important arguments have been presented that any many-particle
quantum system is characterized by some {\it cornerstone} relation 
connecting the single-particle momentum distribution with the in-medium 
pair wave functions. Exactly it is responsible for influence of the 
pairwise interaction on the kinetic energy of quantum gases. The form 
of this relation for a repulsive ground-state uniform Fermi gas has been 
established in the leading-order in the dilution parameter $k_Fa$. It 
is worth noting that if incorporated in the formalism of the $2-$matrix, 
the relation under question allows for getting Schr\"odinger-like 
equations for the in-medium pair wave functions. The simplest variant 
of these equations reduces to the Bethe-Goldstone one. More elaborated 
version demonstrates important and deep parallels with the Galitskii 
consideration. Thus, one can expect that the failure of the 
pseudopotential approach with $E_{\rm kin}$ and $E_{\rm int}$ and, 
also, with the spatial particle correlations occurs somewhere during its 
second step when the effective interaction extracted from the Galitskii 
equation is inserted in the weak-coupling expansions for the physical
quantities.     

This work was supported by the RFBR grant 00-02-17181. The author thanks 
A.Yu. Cherny for the helpful discussions.
 
%--------------------------------------------------------------------

%--------------------------------------------------------------
\appendix
\section{}
\label{app}
In the appendix the situation of a dilute gas of fermions with the 
single-particle spin $\ell/2$ ($\ell >1$ is some integer parameter)
is under investigation. Let particles with some two $z-$projections 
of single-particle spin be present, for example, $\ell/2$ and 
$\ell/2-\ell'$, where $\ell'$ is integer and $\ell'\leq \ell$. And
let the density of the fermions with the projection $\ell/2$ is 
$n_{\ell/2}$, whereas $n_{\ell/2-\ell'}$ corresponds to $\ell/2-
\ell'$. This situation is of special interest in view of the proposal 
of Ref.~\cite{stoof} where a spin pattern like this was suggested 
to strengthen the pairwise-interaction effects in a magnetically 
trapped vapour of $^6Li-$fermions. In the case under consideration 
it turned out to be more convenient to compose a set of the 
eigenfunctions for the $2-$matrix in a way slightly different with 
respect to that of Section~\ref{sec4}. The difference is that we now 
adopt $\nu=\{{\bf q}, {\bf Q}, S_Z, \delta_{S_Z}\}$, where $S_Z$ stands 
for eigenvalues of $\hat{S_Z}$, the $Z-$projection of the total pair 
spin, while $\delta_{S_Z}$ does not correspond to some important 
physical characteristics but simply enumerates degenerate states 
(for more detail, see formulae below). So, here we do not care about 
the total pair spin, in order to simplify all calculations. This choice 
is quite under the general rules of the quantum mechanics together with 
that of Section~\ref{sec4} (see the discussion concerning the 
eigenstates of the $2-$matrix in this section). Equation~(\ref{factor}) 
in the situation of interest is of the form 
\begin{equation}
\varphi_{\nu}({\bf r},\sigma_1,\sigma_2)=
\varphi^{(\delta)}_{{\bf q},{\bf Q}}({\bf r})\,\chi^{(\delta)}_{S_Z}
\!(\sigma_1,\sigma_2),
\label{factor1}
\end{equation}
where the spin variable $\sigma_i\;(i=1,2)$ takes the values $\ell/2$
and $\ell/2-\ell'$. In Eq.~(\ref{factor1}) the condensed notation 
$\delta=\delta_{S_Z}$ is introduced, and $\delta$ is taken as a 
superscript to stress the fact that $\delta$ is here some auxiliary 
quantity rather than an important physical characteristics. The 
eigenstates for $\hat{S}_Z$ are now chosen as follows:
{\setlength\arraycolsep{0pt}
\begin{eqnarray}
&&\bigl|\chi^{(\delta=1)}_{S_Z=\ell}\bigr\rangle=\bigl|
\frac{\ell}{2}\bigr\rangle\bigl|\frac{\ell}{2}\bigr\rangle,
\nonumber\\[1mm]
&&\bigl|\chi^{(\delta=1)}_{S_Z=\ell-\ell'}\bigr\rangle=
\frac{1}{\sqrt{2}}\biggl(\bigl|\frac{\ell}{2}\bigr\rangle\bigl|
\frac{\ell}{2}-
\ell'\bigr\rangle+\bigl|\frac{\ell}{2}-\ell'\bigr\rangle
\bigl|\frac{\ell}{2}\bigr\rangle\biggr),\nonumber\\ [1mm]
&&\bigl|\chi^{(\delta=2)}_{S_Z=\ell-\ell'}\bigr\rangle=
\frac{1}{\sqrt{2}}
\biggl(\bigl|\frac{\ell}{2}
\bigr\rangle\bigl|\frac{\ell}{2}-\ell'\bigr\rangle-
\bigl|\frac{\ell}{2}-\ell'\bigr\rangle
\bigl|\frac{\ell}{2}\bigr\rangle\biggr),\nonumber\\[1mm]
&&\bigl|\chi^{(\delta=1)}_{S_Z=\ell-2\ell'}\bigr\rangle=\bigl|
\frac{\ell}{2}-\ell'\bigr\rangle
\bigl|\frac{\ell}{2}-\ell'\bigr\rangle.
\label{spinstate}
\end{eqnarray}}As it is seen, there is no degeneracy for 
$S_Z=\ell$ and $S_Z=\ell-2\ell'$ when the only variant 
$\delta_{S_Z}=1$ is involved. While for $S_Z=\ell-\ell'$ two 
possible eigenstates of $\hat{S}_Z$ are present: 
$\delta_{\ell-\ell'}=1,2$. From Eq.~(\ref{spinstate}) it follows 
that 
\begin{equation}
\chi^{(1)}_{S_Z}(\sigma_1,\sigma_2)=
\chi^{(1)}_{S_Z}(\sigma_2,\sigma_1),
\label{d1A}
\end{equation}
and, due to the Fermi statistics, we find
\begin{equation}
\varphi^{(1)}_{{\bf q},{\bf Q}}({\bf r})=
-\varphi^{(1)}_{{\bf q},{\bf Q}}(-{\bf r})=
-\varphi^{(1)}_{-{\bf q},{\bf Q}}({\bf r}),
\label{d1B}
\end{equation}
the wave function $\varphi^{(1)}_{{\bf q},{\bf Q}}({\bf r})$ 
approaching $\sqrt{2}\sin({\bf q}{\bf r})$ when $r \to \infty$. 
On the contrary, for $\delta=2$ one gets
\begin{equation}
\chi^{(2)}_{S_Z}(\sigma_1,\sigma_2)=
-\chi^{(2)}_{S_Z}(\sigma_2,\sigma_1),
\label{d2A}
\end{equation}
which results in
\begin{equation}
\varphi^{(2)}_{{\bf q},{\bf Q}}({\bf r})=
\varphi^{(2)}_{{\bf q},{\bf Q}}(-{\bf r})=
\varphi^{(2)}_{-{\bf q},{\bf Q}}({\bf r}),
\label{d2B}
\end{equation}
where $\varphi^{(2)}_{{\bf q},{\bf Q}}({\bf r})$ tends to $\sqrt{2}
\cos({\bf q}{\bf r})$ for $r \to \infty$.

Now, one has everything at his disposal, to express the pair 
correlation function $F_2$and interaction energy $E_{\rm int}$
in terms of PWF. Taken in conjunction with Eqs.~(\ref{2mB}), 
(\ref{pwfA}) and (\ref{pwfB}), Eq.~(\ref{factor1}) in the 
thermodynamic limit yields
{\setlength\arraycolsep{0pt}
\begin{eqnarray}
F_2(x_1,x_2;x_1,x_2)=&&\!\!\!\sum\limits_{S_Z,\delta}
\int \frac{d^3\!{\rm q}\,d^3\!{\rm Q}}{(2\pi)^6}\rho^{(\delta)}_{S_Z}
\!({\bf q},{\bf Q})\nonumber\\[1mm]
&\times&\;|\varphi^{(\delta)}_{{\bf q},{\bf Q}}
({\bf r})|^2\;|\chi^{(\delta)}_{S_Z}\!(\sigma_1,\sigma_2)|^2.
\label{2mnew}
\end{eqnarray}}Further, inserting Eq.~(\ref{2mnew}) in 
Eq.~(\ref{eint3}) and keeping in mind  
$$\sum_{\sigma_1,\sigma_2}
|\chi^{(\delta)}_{S_Z}\!(\sigma_1,\sigma_2)|^2=1,
$$ 
one can arrive at
{\setlength\arraycolsep{0pt}
\begin{eqnarray}
E_{\rm int}/V&=&\frac{1}{2}\!\int\! d^3r\; V(r)
\nonumber\\&\times&\!\sum\limits_{S_Z,\delta}
\int\!\frac{d^3\!{\rm q}\,d^3\!{\rm Q}}{(2\pi)^6}\; 
\rho^{(\delta)}_{S_Z}\!({\bf q},{\bf Q})\;|
\varphi^{(\delta)}_{{\bf q},{\bf Q}}({\bf r})|^2.
\label{eintnewA}
\end{eqnarray}}The low-momentum approximation (for more detail, see 
Eq.~(\ref{lowp})) makes it possible to extremely simplify this 
expression if the dilution limit is of interest. So, for 
$n_{\ell/2},n_{\ell/2-\ell'}\to 0$ one gets 
\begin{equation}
E_{\rm int}/V\simeq\frac{\eta^{(2)}_{\ell-\ell'}}{2}\int d^3r V(r) 
|\varphi^{(2)}({\bf r})|^2,
\label{eintnewB}
\end{equation}
where
\begin{equation}
\eta^{(\delta)}_{S_Z}\!=\!\!\int\!\frac{d^3\!{\rm q}\,d^3\!
{\rm Q}}{(2\pi)^6}\;\rho^{(\delta)}_{S_Z}({\bf q},{\bf Q})
\label{etanewA}
\end{equation}
and
\begin{equation}
\varphi^{(\delta)}({\bf r})=\lim\limits_{{\rm q,Q}\to 0}
\varphi^{(\delta)}_{{\bf q},{\bf Q}}({\bf r}).
\label{vard}
\end{equation}
Only the terms corresponding to $\delta=2$ make contribution to the
interaction energy (\ref{eintnewB}) in the dilution limit because 
$\varphi^{(1)}({\bf r})$ is exactly equal to zero as it is seen from
Eqs.~(\ref{d1B}) and (\ref{vard}). Following the arguments of 
Section~\ref{sec4} (see the discussion just after Eq.~(\ref{eint5})) 
one can rewrite Eq.~(\ref{eintnewB}) in the form
\begin{equation}
E_{\rm int}/V\simeq
         \eta^{(2)}_{\ell-\ell'}\int d^3r V(r) \varphi^2(r)
\label{eintnewC}
\end{equation}
with $\varphi(r)$ obeying Eq.~(\ref{schr}). 

To have an idea about $\eta^{(2)}_{\ell-\ell'}$, let us turn to 
Eq.~(\ref{etaC}). Inserting Eq.~(\ref{2mnew}) in Eq.~(\ref{etaC}), 
one can find
\begin{equation}
\sum\limits_{S_Z,\delta}
\int\frac{d^3\!{\rm q}\,d^3\!{\rm Q}}{(2\pi)^6}\rho^{(\delta)}_{S_Z}
\!({\bf q},{\bf Q})\,|\chi^{(\delta)}_{S_Z}\!(\sigma_1,
\sigma_2)|^2=n_{\sigma_1}n_{\sigma_2}
\label{2mnewC}
\end{equation}
provided that the necessary normalization condition $(1/V)\int_V d^3r 
|\varphi^{(\delta)}_{{\bf q},{\bf Q}}({\bf r})|^2=1$ is taken into 
account (see Eqs.~(\ref{pwfB}) and (\ref{factor1})). From 
Eqs.~(\ref{spinstate}) and (\ref{2mnewC}) it follows that
{\setlength\arraycolsep{0pt}
\begin{eqnarray}
\frac{1}{2}\int\frac{d^3\!{\rm q}\,d^3\!{\rm Q}}{(2\pi)^6}
\Bigl[\rho^{(1)}_{\ell-\ell'}\!({\bf q},{\bf Q})+
      \rho^{(2)}_{\ell-\ell'}\!&&({\bf q},{\bf Q})\Bigr]=\nonumber\\
&&=n_{\ell/2}n_{\ell/2-\ell'}.
\label{etanewB}
\end{eqnarray}}It is natural to expect that $\rho^{(1)}_{\ell-
\ell'}\!({\bf q},{\bf Q})=\rho^{(2)}_{\ell-\ell'}\!({\bf q},{\bf Q})$. 
Hence, from Eqs.~(\ref{etanewA}) and (\ref{etanewB}) one can find
\begin{equation}
\eta^{(\delta)}_{\ell-\ell'}=n_{\ell/2}n_{\ell/2-\ell'}.
\label{etanewC}
\end{equation}     

Finally, using Eqs.~(\ref{theorem}), (\ref{der}), (\ref{eintnewC}) 
and (\ref{etanewC}), for $n_{\ell/2},\,
n_{\ell/2-\ell'} \to 0$ one gets
\begin{equation}
E_{\rm int}/V\simeq[4\pi \hbar^2 (a-b)/m]\,n_{\ell/2} n_{\ell/2-\ell'}
\label{eintnewD}
\end{equation}
and, hence,
\begin{eqnarray}
&&E/V\simeq(4\pi \hbar^2 a/m)\, n_{\ell/2} n_{\ell/2-\ell'},
\label{enew}\\
&&E_{\rm kin}/V\simeq(4\pi \hbar^2 b/m)\,n_{\ell/2} n_{\ell/2-\ell'}.
\label{ekinnew}
\end{eqnarray}
So, as it has been proposed in Ref.~\cite{stoof} and also seen from 
Eqs.~(\ref{eintnewD})-(\ref{ekinnew}), trapping of fermions with two 
different $z-$projections of the single-particle spin allows one to 
operate with a system where the effects of the pairwise interaction 
play a more significant role as compared to the situation of extremely 
weak interacting system of one fermion species. This makes it possible 
to take advantage of a large negative triplet scattering length in 
experiments with a degenerate Fermi gas. Indeed, the stronger 
interaction, the higher BCS-transition temperature~\cite{stoof}.
And the latter significantly simplifies the experimental program of 
searching for the BCS phase transition, especially in the regime $|a| 
\to \infty$.
 

\begin{thebibliography}{99}
\bibitem{ander} M. H. Anderson, J. R. Ensher, M. R. Matthews, C.
E. Wieman, and E. A. Cornell, Science {\bf 269}, 198 (1995).
\bibitem{davis} K. B. Davis et al., Phys. Rev. Lett. {\bf 75},
3969 (1995).
\bibitem{brad} C. C. Bradley, C. A. Sackett, J. J. Tollett, and R.
G. Hulet, Phys. Rev. Lett. {\bf 75}, 1687 (1995).
\bibitem{gp} E. P. Gross, Nuovo Cimento {\bf 20}, 454 (1961);
L. P. Pitaevskii, Zh. Eksp. Teor. Fiz. {\bf 40}, 646
(1961)[Sov. Phys. JETP {\bf 13}, 451 (1961)]
\bibitem{hau} L. Hau et al., Phys. Rev. A {\bf 58}, R54 (1998).
\bibitem{hol} M. J. Holland, D. Jin, M. L. Chiofalo, and J.
Cooper, Phys. Rev. Lett. {\bf 78}, 3801 (1997).
\bibitem{liebA} E. H. Lieb, R. Seiringer, and J. Yn\-gva\-son,
Phys. Rev. A {\bf 61}, 043602 (2000).
\bibitem{liebB}E. H. Lieb, R. Seiringer, J. Yn\-gva\-son,
{\it Su\-per\-flui\-dity in dilute trapped Bose gases}
arXiv: cond-mat/ 0205570.
\bibitem{andrews}  M. R. Andrews et al., Science {\bf 275}, 637
(1997).
\bibitem{kett} S. Inouye et al., Phys. Rev. Lett. {\bf 87}, 080402 
(2001); F. Chevy et al., Phys. Rev. A {\bf 64}, 031601 (2001).
\bibitem{cirac} M. Naraschewski, H. Wallis, A. Schenzle, J. I. Cirac,
P. Zoller, Phys. Rev. A {\bf 54}, 2185 (1996).
\bibitem{devr} J. Tempere, J. T. Devreese, Solid State Comm. {\bf 108},
993 (1998).
\bibitem{pit} F. Dalfovo, S. Giorgini, L. P. Pitaevskii, S.
Stringari, Rev. Mod. Phys. {\bf 71}, 463 (1999).
\bibitem{legett} A. J. Legett, Rev. Mod. Phys. {\bf 73}, 307
(2001).
\bibitem{marco} B. De Marco and D. S. Jin, Science {\bf 285}, 1703
(1999).
\bibitem{schreck} F. Schreck et al., Phys. Rev. A {\bf 64},
011402R (2001).
\bibitem{hadzibabic} Z. Hadzibabic et al., Phys. Rev. Lett. {\bf
88}, 160401 (2002).
\bibitem{roati} G. Roati, F. Riboli, G. Modugno, and M. Inguscio,
Phys. Rev. Lett. {\bf 89}, 150403 (2002).
\bibitem{stoof} H. T. C. Stoof et al., Phys. Rev. Lett. {\bf 76},
10 (1996).
\bibitem{bourdel} T. Bourdel et al., arXiv:cond-mat/0303079, 5 Mar 
2003.
\bibitem{ohara} K. M. O'Hara, S. L. Hemmer, M. E. Gehm, S. R. Granade,
J. E. Thomas, Science {\bf 298}, 2179 (2002).
\bibitem{combescot} R. Combescot, Phys. Rev. Lett. {\bf 87},
080403 (2001).
\bibitem{houbiers} M. Houbiers and H. T. C. Stoof,
Phys. Rev. {\bf A}, 1556 (1999).
\bibitem{vichi} L. Vichi and S. Stringary, Phys. Rev. A {\bf 60}, 
4734 (1999).
\bibitem{shanA} A. Yu. Cherny and A. A. Shanenko, Eur. Phys. J. B {\bf
19}, 555 (2001).
\bibitem{shanB} A. Yu. Cherny and A. A. Shanenko, Phys. Rev. E {\bf
62}, 1646 (2000); {\it ibid}, Phys. Rev. E {\bf 64}, 027105 (2001);
{\it ibid}, Phys. Lett. A {\bf 292}, 287 (2002).
\bibitem{note1} All the derived results can easily be generalized to
the case of another spin choice, see Appendix.
\bibitem{hy} K. Huang and C. N. Yang, Phys. Rev. {\bf 105}, 767 
(1957).
\bibitem{ly} T. D. Lee and C. N. Yang, Phys. Rev. {\bf 105}, 1119
(1957).
\bibitem{brue} K. A. Brueckner and K. Sawada, Phys. Rev. {\bf 106},
1117 (1957).
\bibitem{gal} V. Galitskii, Sov. Phys. JETP {\bf 7}, 104 (1958).
\bibitem{liebC} E.H. Lieb, J. Yngvason, Phys. Rev. Lett. {\bf 80}, 
2504 (1998).
\bibitem{gora} D.S. Petrov {\it et al.}, Phys. Rev. Lett. {\bf 84}, 
2551 (2000).
\bibitem{ties} E. Tiesinga {\it et al.}, J. Res. Natl. Inst. Stand.  
Technol. {\bf 101}, 505 (1996).
\bibitem{bog} N. N. Bogoliubov, {\it Lectures on Quantum Statistics},
(Gordon and Breach, New York, 1970), Vol. 2.
\bibitem{note2} Strictly speaking, the total system momentum is nearly 
conserved if the volume is large enough. It exactly becomes a conserved 
quantity only in the thermodynamic limit.
\bibitem{shanC} A. Yu. Cherny ans A. A. Shanenko, Phys. Rev. B 
{\bf 60}, 1276 (1999).
\bibitem{note3} Here we do not treat neglect of the bound 
pair states and the thermodynamic limit as approximations.
\bibitem{note4} We use the following terminology. ``The weakly 
nonideal Fermi gas'' implies that the total energy of the system
is close to that of the ideal Fermi gas. This gas can be weakly 
interacting when the pairwise interaction is weak: $V(r)=\gamma 
\Phi(r)$ and $\gamma \ll 1$~(see Section \ref{sec3}). It can also 
be a strong-coupling Fermi gas, if $\gamma$ is not small or the
interaction kernel $\Phi(r)$ is singular. In the former case
the system is close in its properties to an ideal Fermi gas due 
to a small coupling constant $\gamma$. In the latter situation 
nonideality is weak due to a small fermion density.
\bibitem{fw} A. L. Fetter and J. D. Walecka, 
{\it Quantum Theory of Many-Particle Systems} (McGraw-Hill, New 
York, 1971).
\bibitem{bg} H. A. Bethe and J. Goldstone, Proc. R. Soc. London,
Ser. A {\bf 238}, 551 (1957).
\end{thebibliography}
\end{document}